\documentclass[apj,twocolappendix, numberedappendixn]{emulateapj}
\usepackage[colorlinks,citecolor=blue]{hyperref}
\usepackage{textcomp}
\usepackage{subfigure}
\usepackage{verbatim}
\usepackage{bm}
\usepackage{hyperref}
\usepackage{amsmath}
\usepackage{graphicx}
\usepackage{epstopdf}
\usepackage{amssymb}
\usepackage{extarrows}
\usepackage{color}
\usepackage{CJK}
\usepackage{cancel}
\usepackage{ulem}

\newcommand{\ba}{\begin{eqnarray}}
\newcommand{\ea}{\end{eqnarray}}
\newcommand{\be}{\begin{equation}}
\newcommand{\ee}{\end{equation}}
\newcommand{\gr}{\mathrm{GR}}
\newcommand{\gw}{\mathrm{GW}}

\newcommand{\m}{\mathrm{max}}
\newcommand{\mi}{\mathrm{min}}

\newcommand{\oct}{\mathrm{oct}}

\newcommand{\au}{\mathrm{AU}}

\newcommand{\IN}{\mathrm{in}}
\newcommand{\OUT}{\mathrm{out}}

\newcommand{\lk}{\mathrm{LK}}

\newcommand{\eff}{\mathrm{eff}}
\newcommand{\f}{\mathrm{f}}
\newcommand{\SL}{\mathrm{sl}}
\newcommand{\li}{\mathrm{lim}}

\def\e1{e_1^2}

\begin{document}
\title{Black Hole and Neutron Star Binary Mergers in Triple Systems: \\II. Merger Eccentricity and Spin-Orbit Misalignment}
\author{Bin Liu$^{1}$, Dong Lai$^{1}$, Yi-Han Wang$^{2}$}
\affil{$^{1}$ Cornell Center for Astrophysics and Planetary Science, Cornell University, Ithaca, NY 14853, USA\\
$^{2}$ Department of Physics and Astronomy, Stony Brook University, Stony Brook, NY 11794-3800, USA
}

\begin{abstract}
We study the dynamical signatures of black hole (BH) and neutron star
(NS) binary mergers via Lidov-Kozai oscillations induced by tertiary
companions in hierarchical triple systems. For each type of binaries
(BH-BH and BH-NS), we explore a wide range of binary/triple
parameters that lead to binary mergers, and
determine the distributions of merger time $T_{\rm m}$, eccentricity
($e_\mathrm{m}$) and spin-orbit misalignment angle ($\theta_\SL^\f$)
when the binary enters the LIGO/VIRGO band (10~Hz). We use the
double-averaged (over both orbits) and single-averaged (over the inner
orbit) secular equations, as well as N-body integration, to evolve
systems with different hierarchy levels, including the leading-order
post-Newtonian effect, de-Sitter spin-orbit coupling and gravitational
radiation. We find that for merging BH-BH binaries with comparable
masses, about $7\%$ have $e_\mathrm{m}>0.1$ and $0.7\%$ have
$e_{\rm m}>0.9$.  The majority of the mergers have significant
eccentricities in the LISA band. The BH spin evolution and the final
spin-orbit misalignment $\theta_\SL^\f$ are correlated with the
orbital evolution and $e_{\rm m}$. Mergers
with negligible $e_{\rm m}$ ($\lesssim 10^{-3}$) have a distribution
of $\theta_\SL^\f$ that peaks around $90^\circ$ (and thus favoring a
projected binary spin parameter $\chi_{\rm eff}\sim 0$), while mergers
with larger $e_{\rm m}$ have a more isotropic spin-orbit
misalignments. For typical BH-NS binaries, strong
octuple effects lead to more mergers with non-negligible
$e_\mathrm{m}$ (with $\sim 18\%$ of the mergers having
$e_\mathrm{m}>0.1$ and $2.5\%$ having $e_\mathrm{m}>0.9$), and the
final BH spin axis tends to be randomly orientated.
Measurements or constraints on eccentric mergers and $\theta_\SL^\f$
from LIGO/VIRGO and LISA would provide useful diagnostics on the dynamical
formation of merging BH or NS binaries in triples.
The recently detected BH merger events may implicate
such dynamical formation channel.
\end{abstract}


\section{Introduction}

Recent studies \citep[e.g.,][]{Miller 2002,Wen 2003,Thompson,Antonini 2012,Silsbee and Tremaine 2017,
Antonini 2017,Liu-ApJL,Liu-ApJ,Hoang 2017,Rodriguez Spin}
have suggested that tertiary-induced merger via Lidov-Kozai (LK) oscillations \citep{Lidov,Kozai,Naoz 2016} may play a
significant role in producing
black-hole (BH) binaries detected by the LIGO/VIRGO collaboration
\citep[e.g.,][]{Abbott 2016a,Abbott 2016b,Abbott 2017a,Abbott 2017b,Abbott 2017c,Abbott 2017d,Abbott 2018a,Abbott 2018b}.
Merging BH and
neutron star (NS) binaries can be formed efficiently in triple systems
with the aid of a tertiary body that moves on an inclined (outer) orbit relative to the orbit
of the inner (BH or NS) binary.
The efficiency of the merger can be further enhanced when the triple is part of a quadruple
system \citep[e.g., when the tertiary component is itself a binary; see][]{Fang 2018,Liu-Quad,Zevin Quadruple,Fragione Quadruple} or
more generally, when the
outer orbit experiences quasi-periodic external forcing \citep[e.g.,][]{Hamers and Lai 2017,Petrovich 2017,Fragione triples}.

Given the expected large number of BH mergers to be detected by LIGO/VIRGO in the
coming years, it will be important to distinguish tertiary-induced mergers
from other dynamical BH binary formation channels \citep[such as those involving
close encounters in dense stellar clusters; e.g.,][]{Portegies
  2000,Miller 2009,O'Leary 2006, Banerjee 2010,Downing
  2010,Rodriguez 2015,Chatterjee 2017,Samsing 2018} and the more traditional
isolated binary channel \citep[e.g.,][]{Lipunov 1997,Lipunov 2017,Podsiadlowski
  2003,Belczynski 2010,Belczynski 2016,Dominik 2012, Dominik
  2013,Dominik 2015}, as well as the chemically homogeneous evolution
channel \citep[e.g.,][]{Mandel and de Mink 2016,Marchant 2016} and gas-assisted mergers \citep[e.g.,][]{Bartos}.
One possible indicator is the merger eccentricity: It has been noted that dynamical binary-single interactions in dense cluster
\citep[e.g.,][]{Samsing 2017,Rodriguez 2018,Samsing 2018b,Fragione GC IMBH} or in galactic triples
\citep[][]{Silsbee and Tremaine 2017,Antonini 2017,Fragione field triple}
may lead to BH binaries entering the LIGO band with modest or large eccentricities,
although the fraction of such eccentric mergers is highly
uncertain. Another potentially valuable observable is the BH spin,
which carries information on the BH binary formation
history. In particular, through the binary inspiral waveform, the mass-weighted projection of BH spin,
$\chi_{\eff}\equiv(m_1 \boldsymbol{\chi}_1+m_2
\boldsymbol{\chi}_2)/(m_1+m_2)\cdot\hat{\mathbf{L}}$, can be directly measured [here $m_{1,2}$ are the BH masses,
$\boldsymbol{\chi}_{1,2}=c\mathbf{S}_{1,2}/(Gm_{1,2}^2)$ are the
dimensionless BH spins, and $\hat{\mathbf{L}}$ is the unit orbital
angular momentum vector of the BH binary]. While isolated binary
evolution tends to generate approximately aligned BH spins (with
respect to the orbit), and cluster dynamics tends to generate random
spin orientations, recent works \citep[][]{Liu-ApJL,Liu-ApJ,Antonini spin,Rodriguez Spin} have shown that
merging BH binaries produced in hierarchical triples may exhibit rich behaviors
in spin-orbit orientations. For initially close BH binaries (with semimajor axis
$a_0\lesssim 0.2\au$), which may merge by themselves without the aid of
a tertiary companion, modest ($\lesssim 20^\circ$)
spin-orbit misalignments can be produced \citep[e.g.,][]{Liu-ApJL} due
to the perturbation of the tertiary companion.
For wide binaries (with $a_0\gtrsim 10\au$),
a range of final spin-orbit
misalignment angles $\theta_{\rm sl}^{\rm f}$ can be produced as the
merging binary enters the LIGO band \citep[][]{Liu-ApJ}: When
the BHs have comparable masses (the octupole effect is thus negligible), the
distribution of $\theta_{\rm sl}^\f$ is peaked
around $90^\circ$; when the two members of inner binary have highly
unequal masses and the tertiary companion moves on an eccentric orbit,
a more isotropic distribution of final spin axis is produced.
Overall, merging BH binaries produced by Lidov-Kozai oscillations in triples exhibit a unique
distribution of the effective (mass-weighted) spin parameter $\chi_{\rm eff}$.

In this paper, we extend our recent studies \citep[][]{Liu-ApJL,Liu-ApJ}
by exploring a wide range of triple systems. In particular, for a given
merging compact binary with known masses, we examine all
possible triple configurations and parameters that lead to binary mergers
and determine the distributions of various merger properties (merger times,
eccentricities and spin-orbit misalignments).
We consider two sets of binary masses: $(m_1,m_2)=
(30M_\odot,20M_\odot)$ representing a canonical BH-BH binary, and
$(30M_\odot,1.4M_\odot)$ representing a canonical BH-NS binary.
Our previous works focused on fully hierarchical triples, where
either double-averaged (over both the inner and outer orbits) or
single-averaged (over only the inner orbit) secular approximation is
valid. The spin evolution was only studied for systems where the
double-average approximation is valid [this is also the case for the
study by \cite{Antonini spin} and \cite{Rodriguez Spin}]. In this paper we examine systems with
various levels of hierarchy, using both double-averaged and
single-averaged secular equations, as well as direct N-body
integrations to evolve the systems.
This allows us to determine reliably the fraction of merging
binaries that enter the LIGO band with appreciable eccentricities.
In addition, unlike previous works, we consider systems where the initial BH spin
and orbital axes are not perfectly aligned and we determine the distribution of the
``final'' spin-orbit misalignment angles $\theta_\SL^\f$.

Our paper is organized as follows. In Section \ref{sec 2},
we introduce three approaches to evolve the
triple systems with different levels of approximation,
including the de-Sitter spin-orbit coupling effect.
In Section \ref{sec 3}, we perform a large set of numerical
integrations, focusing on two types of stellar mass binaries (BH-BH
and BH-NS), with imperfectly aligned initial spin axes.  We compute the
distributions of binary eccentricities and spin-orbit misalignments
for systems that evolve into the LIGO band.
We summarize our main results in Section \ref{sec 4}.

\section{Three Approaches for Triple Evolution with Spin-Orbit coupling}
\label{sec 2}
\subsection{Summary of Parameter Regimes}
\label{sec 2 1}

We consider a hierarchical triple system, composed of
an inner BH binary of masses $m_1$, $m_2$
and a distant companion of mass $m_3$
that moves around the center of mass of the inner bodies.
The reduced mass for the inner binary is $\mu\equiv m_1m_2/m_{12}$, with $m_{12}\equiv m_1+m_2$.
Similarly, the outer binary has $\mu_\OUT\equiv(m_{12}m_3)/m_{123}$ with $m_{123}\equiv m_{12}+m_3$.
The semi-major axes and eccentricities are denoted by $a$, $a_\OUT$ and $e$, $e_\OUT$, respectively.
Therefore, the orbital angular momenta of two orbits are given by
$\textbf{L}=\mathrm{L}\hat{\textbf{L}}=\mu\sqrt{G m_{12}a(1-e^2)}\,\hat{\textbf{L}}$
and $\textbf{L}_\OUT=\mathrm{L}_\OUT\hat{\textbf{L}}_\OUT=\mu_\OUT\sqrt{G m_{123}a_\OUT(1-e_\OUT^2)}\,\hat{\textbf{L}}_\OUT$.
We define the mutual inclination between $\textbf{L}$ and $\textbf{L}_\OUT$ (inner and outer orbits) as $I$.

To study the evolution of the inner binary under
the influence of the tertiary companion, we use three approaches: the
double-averaged (averaging over both the inner and outer orbits),
and single-averaged (only averaging over the inner orbital period) secular equations of motion,
as well as the direct N-body integrations [see Section 2.1 of \cite{Liu-ApJ} for details].
In the orbital evolution, we include the contributions from the external
companion that generate LK oscillations up to the octupole level of approximation, the post-Newtonian (PN) correction due to general
relativity (GR), and the dissipation due to gravitational wave (GW) emission.

The LK mechanism induces the oscillations in the eccentricity
and mutual orbital inclination on the timescale
\be
t_\lk=\frac{1}{n}\frac{m_{12}}{m_3}\bigg(\frac{a_{\OUT,\eff}}{a}\bigg)^3,
\ee
where $n=(G m_{12}/a^3)^{1/2}$ is the mean motion of the inner binary,
and $a_{\OUT,\eff}\equiv a_\OUT\sqrt{1-e^2_\OUT}$ is the effective outer binary separation.

During the LK oscillations, the short-range force effects (such as GR-induced apsidal precession)
play a crucial role in determining the maximum eccentricity $e_\m$ of the inner binary
\citep[e.g.,][]{Fabrycky and Tremaine 2007}.
In the absence of energy dissipation, the evolution of the
triple is governed by two conservation laws:
the total orbital angular momentum and
the total energy of the system.
The analytical expression for $e_\m$ for general hierarchical triples
(arbitrary masses and eccentricities) can be
obtained in the double-averaged secular approximation if the disturbing potential
is truncated to the quadrupole order.
Using the method of \citet{Liu et al 2015} \citep[see also][]{Anderson et al HJ,Anderson et al 2017}, we find
\ba\label{eq:EMAX}
&&\frac{3}{8}\Bigg\{e_0+(j_\mi^2-1)+(5-4j_\mi^2)\\
&&\times\Bigg[1-\frac{\Big((j_\mi^2-1)\eta_\mi+e_0^2\eta_0-2j_0\cos I_0\Big)^2}{4j_\mi^2}\Bigg]\nonumber\\
&&-(1+4e_0^2-5e_0^2\cos^2\omega_0)\sin^2I_0\Bigg\}+\varepsilon_\gr \left(j_0^{-1}-j_\mi^{-1}\right)=0\nonumber,
\ea
where $e_0$, $I_0$ and $\omega_0$ are the initial eccentricity, inclination and longitude of the periapse of the inner binary, respectively,
and we have defined
$j_\mi\equiv\sqrt{1-e_\m^2}$, $j_0\equiv\sqrt{1-e_0^2}$, $\eta_\mi\equiv \mathrm{L}(e=e_\m)/\mathrm{L}_\OUT$, $\eta_0\equiv\mathrm{L}(e=e_0)/\mathrm{L}_\OUT$ and
$\varepsilon_\gr=(3Gm_{12}^2a_{\OUT,\eff}^3)/(c^2a^4m_3)$.
Note that for $e_0=0$, Equation~(\ref{eq:EMAX}) reduces to Equation (24) of \cite{Anderson et al 2017}.
For the general $\mathrm{L}/\mathrm{L}_\OUT$,
the maximum possible $e_\m$ for all values of $I_0$, called
$e_\li$, is given by (assuming $\omega_0=0$)
\ba\label{eq:ELIM}
&&\frac{3}{8}\Bigg\{(j_\li^2-1)\Bigg[\frac{\eta_\mi^2}{4}\Bigg(\frac{4}{5}j_\li^2-1\Bigg)-3\Bigg]\\
&&+\frac{e_0^4}{(j_\li^2-1)}\frac{\eta_0^2}{4}\Bigg(\frac{4}{5}j_\li^2-1\Bigg)\nonumber\\
&&+2e_0^2\Bigg[\frac{\eta_\mi\eta_0}{4}\Bigg(\frac{4}{5}j_\li^2-1\Bigg)+1\Bigg]\Bigg\}
+\varepsilon_\gr \left(j_0^{-1}-j_\li^{-1}\right)=0\nonumber,
\ea
where $j_\li\equiv\sqrt{1-e_\li^2}$.

For systems with $m_1\neq m_2$ and $e_\OUT\neq 0$,
the octupole effect may become important. The strength of the octupole effect is
characterized by
\be\label{eq:epsilon oct}
\varepsilon_\oct\equiv\frac{m_1-m_2}{m_1+m_2}\frac{a}{a_\OUT}\frac{e_\OUT}{1-e_\OUT^2}.
\ee
The octupole terms tend to widen the inclination window for large eccentricity excitation.
However, the analytic expression for $e_\li$ given by
Equation~(\ref{eq:ELIM}) remains valid even when the octupole effect is strong \citep{Liu et al 2015, Munoz 2016,Anderson et al 2017}.

The validity of secular approximation depends on the hierarchy level of the triple system.
For sufficiently hierarchical systems, the angular momenta of the inner and outer binaries exchange periodically over a long timescale
(longer than the companion's orbital period),
while the exchange of energy is negligible.
When the eccentricity variation timescale of the inner binary
is longer than the period of companion's orbit ($P_\OUT$), i.e.,
\be\label{eq:DA}
t_\lk\sqrt{1-e_\m^2}\gtrsim P_\OUT,
\ee
the double-averaged (DA) secular equations are valid.
The full equations of motion can be found in \citet{Liu et al 2015}.

For moderately hierarchical systems, when the eccentricity evolution timescale at $e\sim e_\m$ lies between the inner orbital
period $P_\IN$ and the outer orbital period $P_\OUT$, i.e.,
\be\label{eq:SA}
P_\IN\lesssim t_\lk\sqrt{1-e_\m^2}\lesssim P_\OUT,
\ee
the DA secular equations break down, but
the single-averaged (SA) secular equations remain valid.
The explicit SA equations of motion are provided in Section 2.1.2 of \citet{Liu-ApJ}.

If the tertiary companion is even closer, the perturbation becomes sufficient strong and
Equation (\ref{eq:SA}) may not be satisfied.
In this situation, the dynamics can only be solved by N-body (NB) integrations.
In this paper, we use a new N-body code developed by Yi-Han Wang based on
the {\tt ARCHAIN} algorithm \citep[][]{archain}.
This algorithm employs a regularized integrator to accurately trace the motion of tight binaries with arbitrarily large mass ratios,
and a chain structure to reduce the round-off errors from close encounters.
The code uses the Bulirsch-Stoer(BS) integrator \citep[e.g.,][]{Stoer} and further reduces the
round-off error by using active error compensation.
Our developing code can be found at
\mbox{\href{https://github.com/YihanWangAstro/Template-SpaceX}{\tt SpaceHub}}.

All calculations in this paper, whether based on DA or SA secular equations, or NB integrations,
include the PN effect of the inner binary (which gives rise to the apsidal advance) and the 2.5 PN effect
(which accounts for gravitational radiation).

\subsection{Spin-Orbit Coupling}
\label{sec 2 2}

To incorporate the spin-orbit coupling effect, we introduce the spin vector
$\textbf{S}_1=\mathrm{S}_1 \hat{\textbf{S}}_1$ (where
$\mathrm{S}_1$ is the magnitude of the spin angular momentum of $m_1$ and
$\hat{\textbf{S}}_1$ is the unit vector).
The de Sitter precession of $\hat{\textbf{S}}_1$ around $\hat{\mathbf{L}}$ (1.5 PN effect)
is governed by \citep[e.g.,][]{Barker}
\be\label{eq:spin}
\frac{d \hat{\textbf{S}}_1}{dt}=\boldsymbol{\Omega}_\mathrm{SL} \times \hat{\textbf{S}}_1.
\ee
Let $\textbf{r}_{1(2)}$ and $\textbf{v}_{1(2)}$
be the position vector and velocity vector of the first and second body of the inner binary, respectively,
and define $\textbf{r}=\textbf{r}_1-\textbf{r}_2$, $r=|\textbf{r}|$, and $\textbf{v}=\textbf{v}_1-\textbf{v}_2$.
The precession rate in Equation (\ref{eq:spin}) is given by
\be\label{eq:desitter rate PN}
~~~\boldsymbol{\Omega}_\mathrm{SL}^\mathrm{PN}=G\bigg(2+\frac{3m_2}{2m_1}\bigg)\frac{\mu\textbf{r}\times\textbf{v}}{c^2r^3}.
\ee
Averaging over the inner orbital period ($P_\IN$), the precession rate becomes
\be\label{eq:desitter rate}
~~~\boldsymbol{\Omega}_\mathrm{SL}^{(\mathrm{AV})}=\frac{3 G n (m_{2}+\mu/3)}{2 c^2 a (1-e^2)}\hat{\textbf{L}}
=\Omega_\mathrm{SL}^{(\mathrm{AV})}\hat{\textbf{L}}.
\ee
Similar equations apply to $\textbf{S}_2$.
Note that the back-reaction torques from $\hat{\textbf{S}}_1$ on $\hat{\mathbf{L}}$
is usually negligible since $S_1\ll L$, and the spin-spin coupling (2 PN correction) is always
negligible until the very last merging stage.
Both are ignored in our calculations.
In addition, the de Sitter precession of $\hat{\mathbf{S}}_1$ induced by the tertiary companion is neglected as well.
Thus, the DA/SA secular equations combined with Equation (\ref{eq:desitter rate}), or
N-body integration with Equation (\ref{eq:desitter rate PN}),
completely determine the orbital and spin evolution of merging BH binaries in triples.

\subsection{Some Examples}
\label{sec 2 3}

\begin{figure}
\centering
\begin{tabular}{cccc}
\includegraphics[width=9cm]{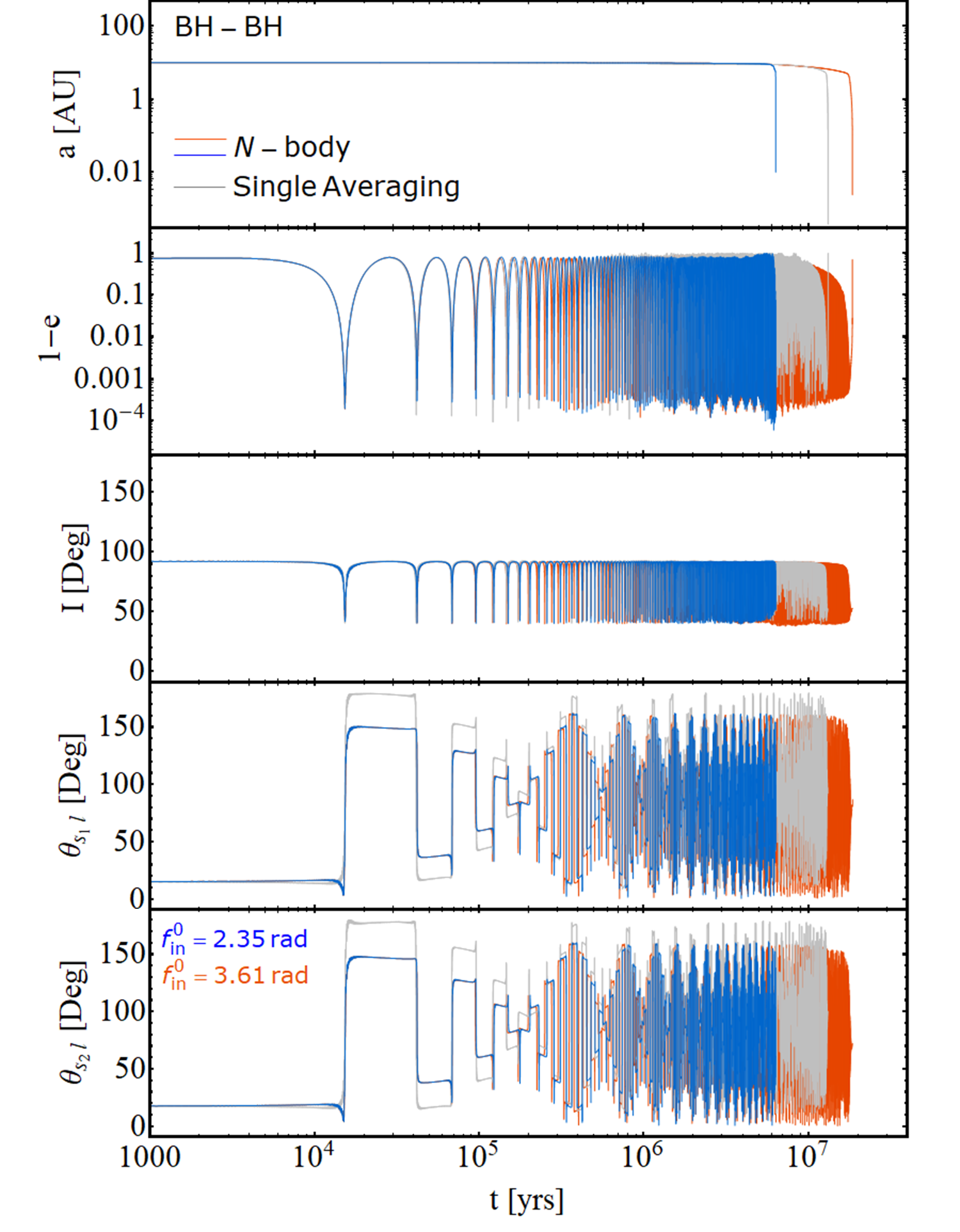}
\end{tabular}
\caption{Sample orbital and spin evolution of a BH binary system with a tertiary companion.
The three top panels show the semi-major axis, eccentricity
and inclination (relative to ${\hat{\bf L}}_\OUT$) of the inner BH binary, and the bottom panel show
the spin-orbit misalignment angle $\theta_\SL$ (the angle between ${\bf S}_1$ and ${\bf L}$).
The system parameters are: $m_1=30M_\odot$, $m_2=20M_\odot$, $m_3=30M_\odot$,
$a_0=10\au$, $a_\OUT=200\au$, $e_0=0.3$, $e_\OUT=0.1$,
$I_0=91.85^\circ$, and $\theta^0_\mathrm{s_1l}=15.13^\circ$, $\theta^0_\mathrm{s_2l}=17.64^\circ$.
The initial longitudes of periapse of the inner and outer orbits (i.e., the angle between $\bf e$ and line of the ascending node of the two orbits) are
$\omega_{\IN,0}=340^\circ$ and $\omega_{\OUT,0}=113^\circ$. Different colors denote
two integration methods. In our calculations,
the initial true anomaly ($f^0_\OUT$) of the outer binary is set to be $\pi$; in the N-body calculations,
the initial true anomaly ($f_\IN^0$) of the inner orbit is set to $2.35$ rad (blue) and $3.61$ rad (red).
}
\label{fig:evolution1}
\end{figure}

To calibrate our different approaches,
Figure \ref{fig:evolution1} shows an example of the orbital and spin evolution of a BH binary
with an inclined companion, obtained using N-body integration and SA secular equations.
The parameters of the system (given in the figure caption) satisfy the SA criterion (Equation \ref{eq:SA}).
We see that
the SA equations succeed in resolving the ``correct" orbital evolution,
producing the same period and amplitude of LK cycles as in the N-body calculations.
However, in the N-body calculations, the long term evolution of the binary depends on the initial true anomaly $f_\IN^0$
of the inner orbit, with the merger time depending on $f_\IN^0$. The evolution based on the SA equation yields an ``averaged" merger time.
Note that in this example, the`` residual" eccentricity (i.e., the binary eccentricity when it enters the LIGO band) is
negligible ($e_\mathrm{m}\ll1$) regardless of the integration methods.

The bottom panel of Figure \ref{fig:evolution1} shows that the spin axis $\hat{\textbf{S}}_1$
(initially misaligned with respect to $\hat{\mathbf{L}}$ by $15^\circ$)
experiences large variations during the inner binary evolution. As discussed in \cite{Liu-ApJ}, a useful ``adiabaticity parameter"
characterizing the spin evolution is \citep[see also][]{Dong Science,Storch spin,Anderson et al HJ,Anderson et al 2017}
\be
\mathcal{A}\equiv\bigg|\frac{\Omega_\mathrm{SL}^{(\mathrm{AV})}}{\Omega_\mathrm{L}}\bigg|, ~~~~~~
\mathrm{with}~~~\Omega_\mathrm{L}=\frac{3(1+4e^2)|\sin2I|}{8t_\lk\sqrt{1-e^2}}.
\ee
As the orbit decays, the spin dynamics transitions from the ``weak coupling" regime
($\mathcal{A}\ll1$) to the ``strong coupling" regime ($\mathcal{A}\gg1$).
The spin-orbit misalignment angle
tends to be frozen at a high value near the end of the inspiral.
In this example, all integrations produce large final spin-orbit misalignment angles,
$\theta_\mathrm{s_1l}^\f \simeq81^\circ$, $\theta_\mathrm{s_2l}^\f \simeq77^\circ$ ($f_\IN^0=2.35$ rad)
and $\theta_\mathrm{s_1l}^\f \simeq87^\circ$, $\theta_\mathrm{s_2l}^\f \simeq72^\circ$ ($f_\IN^0=3.61$ rad) for the N-body integration, and
$\theta_\mathrm{s_1l}^\f\simeq\theta_\mathrm{s_2l}^\f\simeq90^\circ$ for the SA secular integration.

\begin{figure}
\centering
\begin{tabular}{cccc}
\includegraphics[width=9cm]{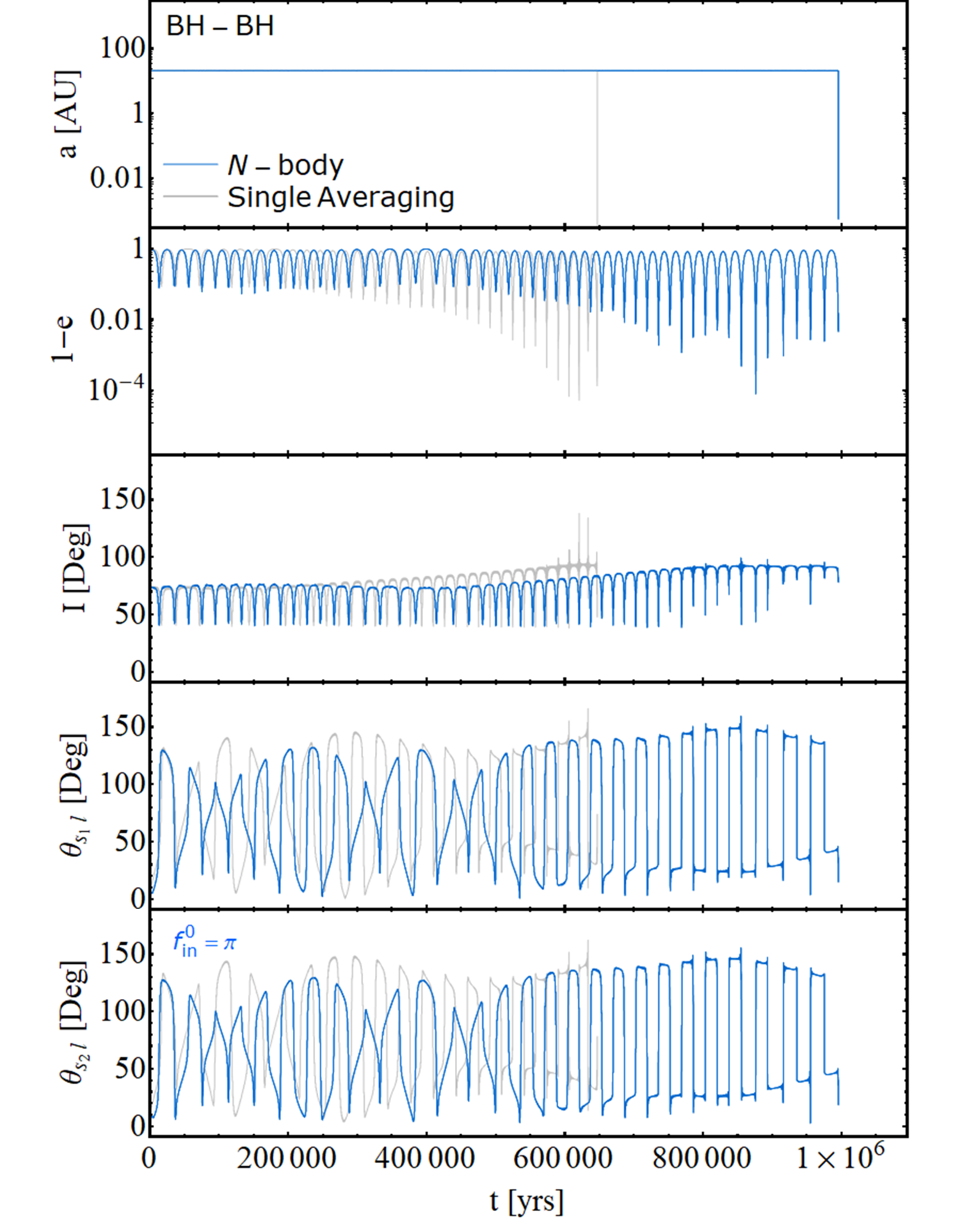}
\end{tabular}
\caption{Another example of the orbital and spin evolution of a BH binary in triple system,
The system parameters are: $m_1=30M_\odot$, $m_2=20M_\odot$, $m_3=49.78M_\odot$,
$a_0=20.80\au$, $a_\OUT=381.39\au$, $e_0=0.017$, $e_\OUT=0.699$,
$I_0=73.60^\circ$, $\omega_{\IN,0}=111.39^\circ$, $\omega_{\OUT,0}=234.54^\circ$,
and $\theta^0_\mathrm{s_1l}=7.9^\circ$, $\theta^0_\mathrm{s_2l}=12.5^\circ$. In this example,
the initial true anomalies of the inner and outer binaries are set to $\pi$.
}
\label{fig:evolution2}
\end{figure}

Figure \ref{fig:evolution2} depicts another example of BH binary
merger, in which the octupole effect is important. 
The system parameters (given in the figure caption) imply that both DA and SA approaches are not accurate.
We see that in the N-body integration, due to the high-efficiency of gravitational radiation at $e_\m$, rapid merger occurs,
accompanied by highly eccentric orbit ($e_\mathrm{m}\simeq0.94$)
when the GW frequency enters the LIGO band.
Because of the rapid orbital decay, the spin vector does not experience large variation and the final spin-orbit misalignment angle
freezes at $\theta_\mathrm{s_1l}^\f \simeq15^\circ$, $\theta_\mathrm{s_2l}^\f \simeq18^\circ$.
However, for the SA secular integration, the residual eccentricity is ``erased" ($e_\mathrm{m}\simeq0.05$) and the final spin-orbit misalignment angle are
$\theta_\mathrm{s_1l}^\f \simeq39^\circ$, $\theta_\mathrm{s_2l}^\f \simeq22^\circ$.

\section{Population Study}
\label{sec 3}
\subsection{Parameter Choice and System Setup}
\label{sec 3 1}

We consider two types of merging binaries in this paper. The first has masses $(m_1, m_2)=(30M_\odot, 20M_\odot)$,
representing typical BH-BH binaries; the second has $(m_1, m_2)=(30M_\odot, 1.4M_\odot)$, representing
BH-NS binaries. For each inner binary, we consider all possible initial binary/triple systems and
parameters that may lead to binary mergers. In particular, the initial semimajor axes
of the inner and outer orbits ($a_0$ and $a_\OUT$) are
chosen from a log-uniform distribution from $10$ AU to $10^4$ AU;
the initial orbital eccentricities are drawn from a uniform distribution ranging from 0 to 1
for both inner and outer binary orbits;
the tertiary companion mass is assigned by assuming a flat distribution in (0, 1) for $m_3/m_{12}$
\citep[e.g.,][]{Sana 2012,Duchene 2013,Kobulnicky 2014};
the binary inclinations are isotropically distributed
(uniform distribution in $\cos I_0$).
Since the velocity kick during the BH formation may introduce
small spin-orbit misalignment, we consider a flat distribution of the initial $\cos\theta_\SL^0$
in the range $(\cos0^\circ,\cos20^\circ)$.

For the triple systems to be dynamically stable,
the ratio of the pericenter distance of the outer orbit to the apocenter distance of the inner
orbit must satisfy \citep[e.g.,][]{Kiseleva}
\be\label{eq:stable}
\frac{a_\OUT(1-e_\OUT)}{a(1+e)}>\frac{3.7}{Q_\OUT}-\frac{2.2}{1+Q_\OUT}+\frac{1.4}{Q_\IN}\frac{Q_\OUT-1}{Q_\OUT+1},
\ee
where $Q_\IN=[\m(m_1,m_2)/\mi(m_1,m_2)]^{1/3}$ and $Q_\OUT=(m_{12}/m_3)^{1/3}$.

For each type of binaries, simulations are preformed for $10^5$ randomly chosen initial conditions.
After extracting the stable triples based on Equation (\ref{eq:stable}), we
identify the parameter regime for each triple according to the criteria of Section \ref{sec 2 1}.
We use $e_\m=e_\li$ in Equations (\ref{eq:DA}) and (\ref{eq:SA}).
Different integration methods (DA/SA/NB) are then applied to systems in different regimes.
This helps us to speed up our calculations and ensures the accuracy of system evolution.

To further increase the efficiency of the parameter survey, we adopt the following ``stopping conditions".
First, the maximum integration time is set to be the minimum of $(10^4~t_\lk, 10~\mathrm{Gyrs})$.
This maximum time is adequate to capture the most of the mergers.
Moreover, we terminate the simulation when the inner binary semimajor axis is reduced to less than $0.5\%$ of the
initial $a_0$ and when the adiabaticity parameter $\mathcal{A}\gg1$.
This is reasonable because in the last phase of the binary merger, the binary dynamics is dominated by GW emission
such that the inner binary is decoupled from the perturbation of the tertiary companion.
The condition $\mathcal{A}\gg1$ ensures that the spin-orbit misalignment angle $\theta_\SL$ reaches its ``final" (constant) value.
Once the full integration is stopped, the subsequent evolution of the inner binary eccentricity and semimajor axis can be obtained
using the analytical formulas of \citet{Peters 1964}.
This allows us to obtain the residual eccentricity $e_\mathrm{m}$
when binary enters LIGO detection band, i.e., when the peak GW frequency \citep[][]{Wen 2003}
\be\label{eq:crita1}
f_\gw^\mathrm{peak} =\frac{(1+e)^{1.1954}}{\pi}\sqrt{\frac{G(m_1+m_2)}{a^3(1-e^2)^3}}
\ee
reaches 10 Hz.

\subsection{Parameter Space for Binary Mergers}
\label{sec 3 2}

\begin{figure}
\centering
\begin{tabular}{cccc}
\includegraphics[width=7.5cm]{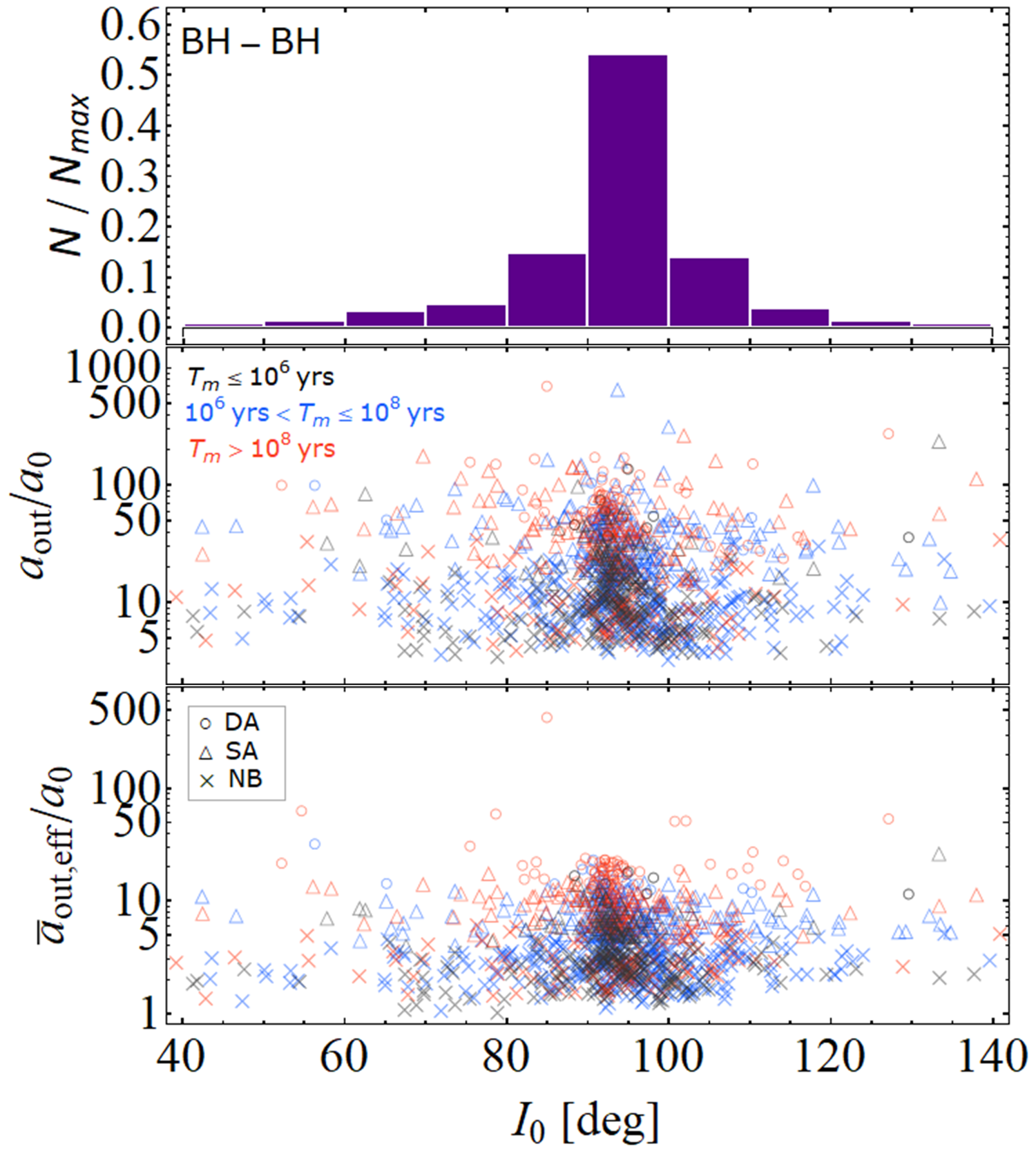}
\end{tabular}
\caption{Parameter space producing BH mergers in triple systems (with $m_1=30M_\odot, m_2=20M_\odot$).
Top panel: the distribution of initial inclination between the inner and outer orbits.
Middle panel: initial binary separation ratio $a_\OUT/a_0$ versus $I_0$. Bottom panel:
ratio between the scaled outer semimajor axis (Equation \ref{eq:aout bar}) and the inner one $\bar{a}_{\OUT,\eff}/a_0$.
Results are separated into three ranges of merger time (colors) and three integration methods (symbols), as labeled:
double-averaged (DA), single-averaged (SA) secular equations and N-body (NB) calculation, which are relevant to
different parameter regimes (see Equations \ref{eq:DA}-\ref{eq:SA}).
}
\label{fig:BH-BH I0}
\end{figure}

What kinds of triple systems can produce binary mergers?
For our canonical BH-BH binaries ($m_1=30M_\odot, m_2=20M_\odot$),
we find 1092 mergers out of 25255 stable systems, with merger fraction $4.3\%$; this includes
114 mergers out of 10243 systems in DA parameter regime (fraction $\simeq 1.1\%$),
373 mergers out of 5785 systems in SA parameter regime (fraction $\simeq 6.5\%$), and 605 mergers out of 9277 systems
from direct NB simulations (fraction $\simeq 6.6\%$).
Figure \ref{fig:BH-BH I0} depicts the parameter space that produces BH mergers, where we plot the initial binary separation ratio
$a_\OUT/a_0$ and $\bar{a}_{\OUT,\eff}/a_0$ versus the inclination ($I_0$) between
the inner and outer orbits. We also plot the distribution of $I_0$ in the top panel. Here,
we introduce the dimensionless scaled semi-major axis
\be\label{eq:aout bar}
\begin{split}
\bar{a}_{\OUT,\eff}\equiv\bigg(\frac{a_\OUT\sqrt{1-e_\OUT^2}}{1\au}\bigg)\bigg(\frac{m_3}{1M_\odot}\bigg)^{-1/3}
\end{split}
\ee
to characterize the ``strength" of the tertiary companion.
The final outcome of each integration is also indicated by the color
(quick, moderate and slow mergers, corresponding to merger time $T_\mathrm{m}\leq10^6 \mathrm{yrs}$,
$10^6 \mathrm{yrs}<T_\mathrm{m}\leq10^8 \mathrm{yrs}$ and $T_\mathrm{m}>10^8 \mathrm{yrs}$, respectively)
and by the symbol (DA, SA and NB integration methods).

\begin{figure}
\centering
\begin{tabular}{cccc}
\includegraphics[width=6.5cm]{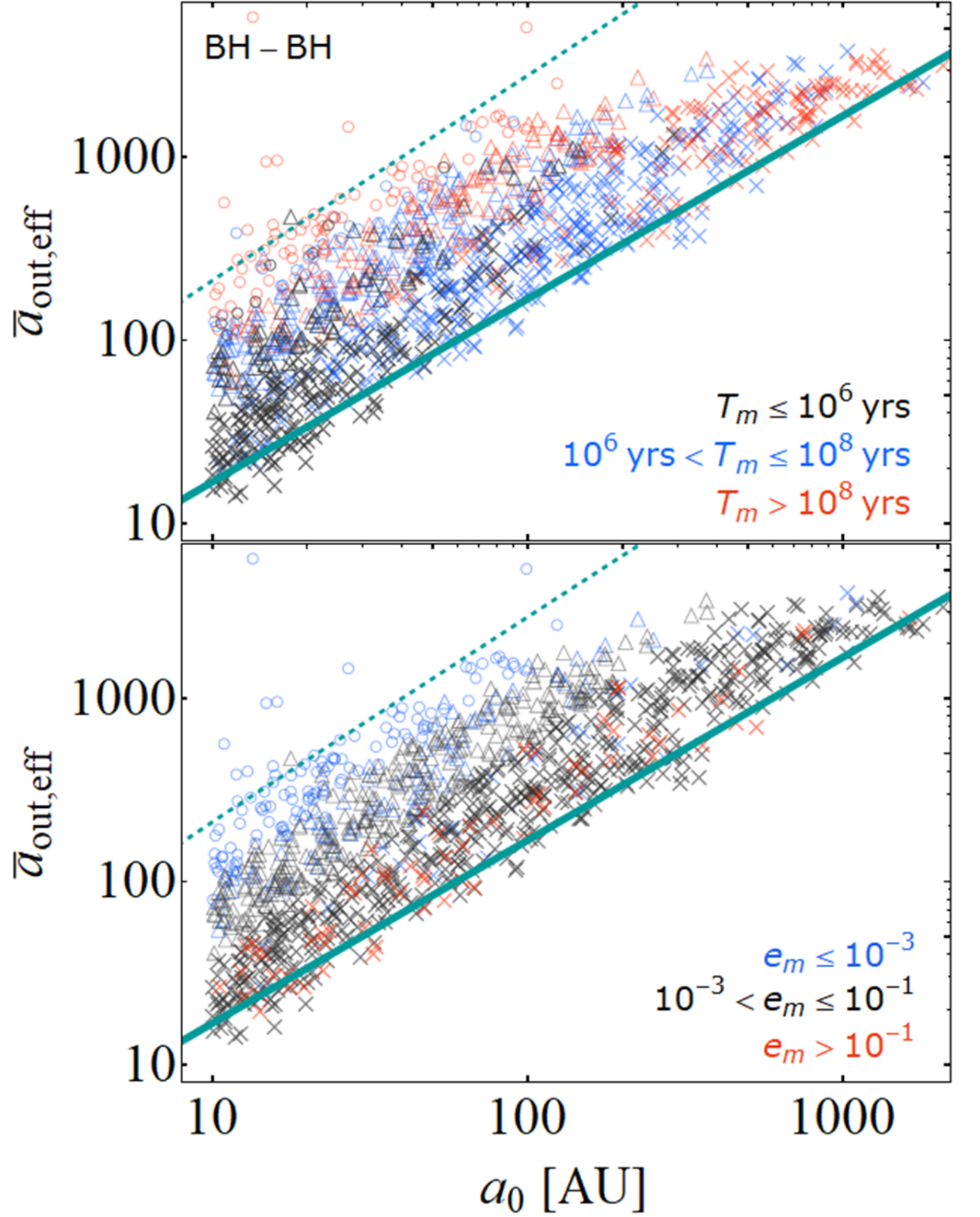}
\end{tabular}
\caption{Parameter space producing BH-BH mergers presented in Figure \ref{fig:BH-BH I0}.
Different merger times and residual eccentricities are color coded as labeled in the upper and lower panels, respectively.
The three types of symbols indicate the mergers achieved by DA, SA and NB integrations (same as Figure \ref{fig:BH-BH I0}).
The solid lines represent the stability criterion (Equation \ref{eq:stable} with $e_0=1$, $e_\OUT=0$ and
$m_3=50M_\odot$). The dashed lines represent the the requirement
for the detectable mergers in LK channel (Equation \ref{eq:fitting formula} with $e_\mathrm{m}$
given by Equation \ref{eq:ELIM} using $e_0=0$ and $\eta_0=0$).
}
\label{fig:BH-BH a0}
\end{figure}

\begin{figure}
\centering
\begin{tabular}{cccc}
\includegraphics[width=7.5cm]{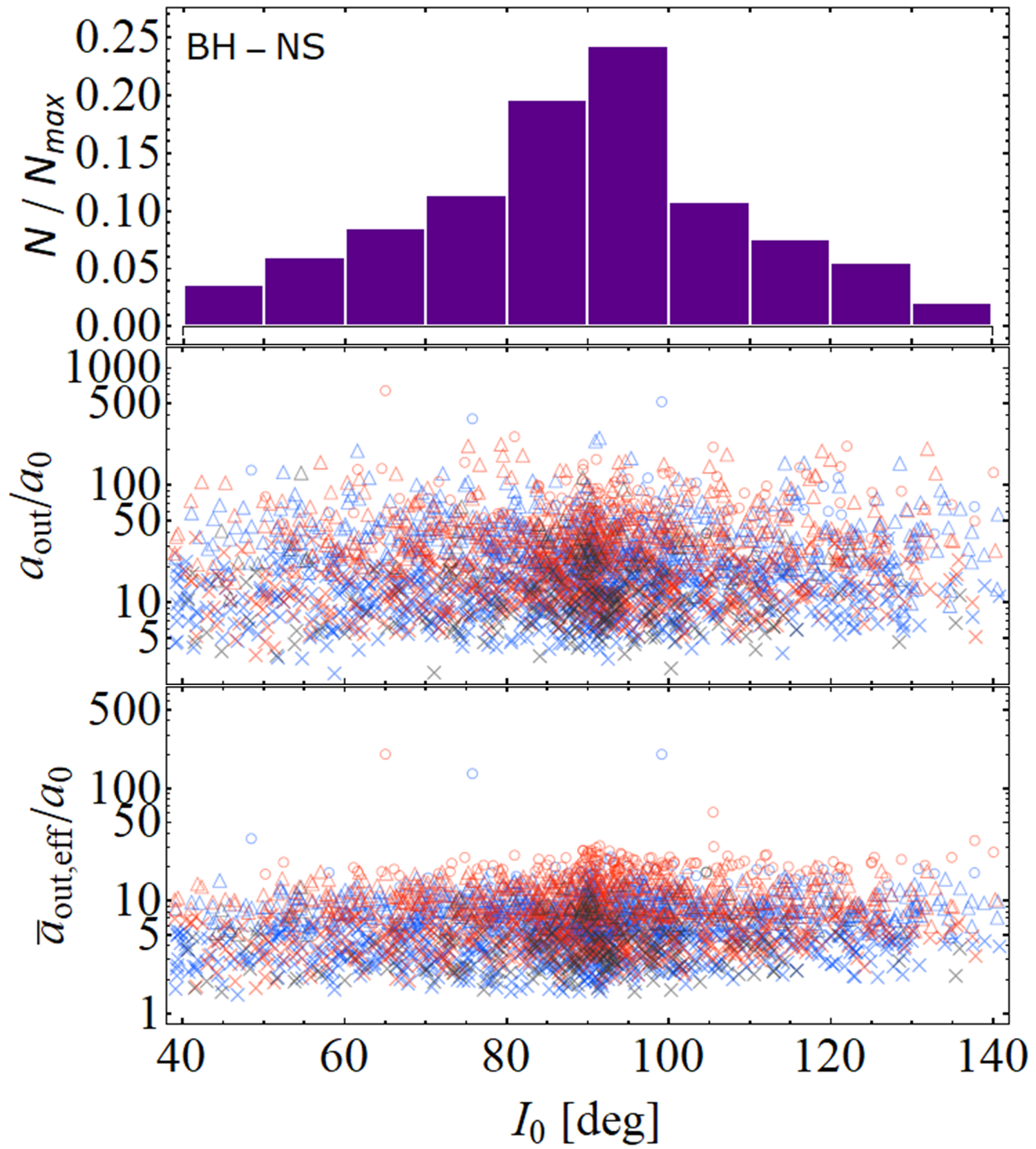}
\end{tabular}
\caption{Same as Figure \ref{fig:BH-BH I0}, but for our canonical BH-NS binaries (with $m_1=30M_\odot, m_2=1.4M_\odot$).
}
\label{fig:BH-NS I0}
\end{figure}

\begin{figure}
\centering
\begin{tabular}{cccc}
\includegraphics[width=6.5cm]{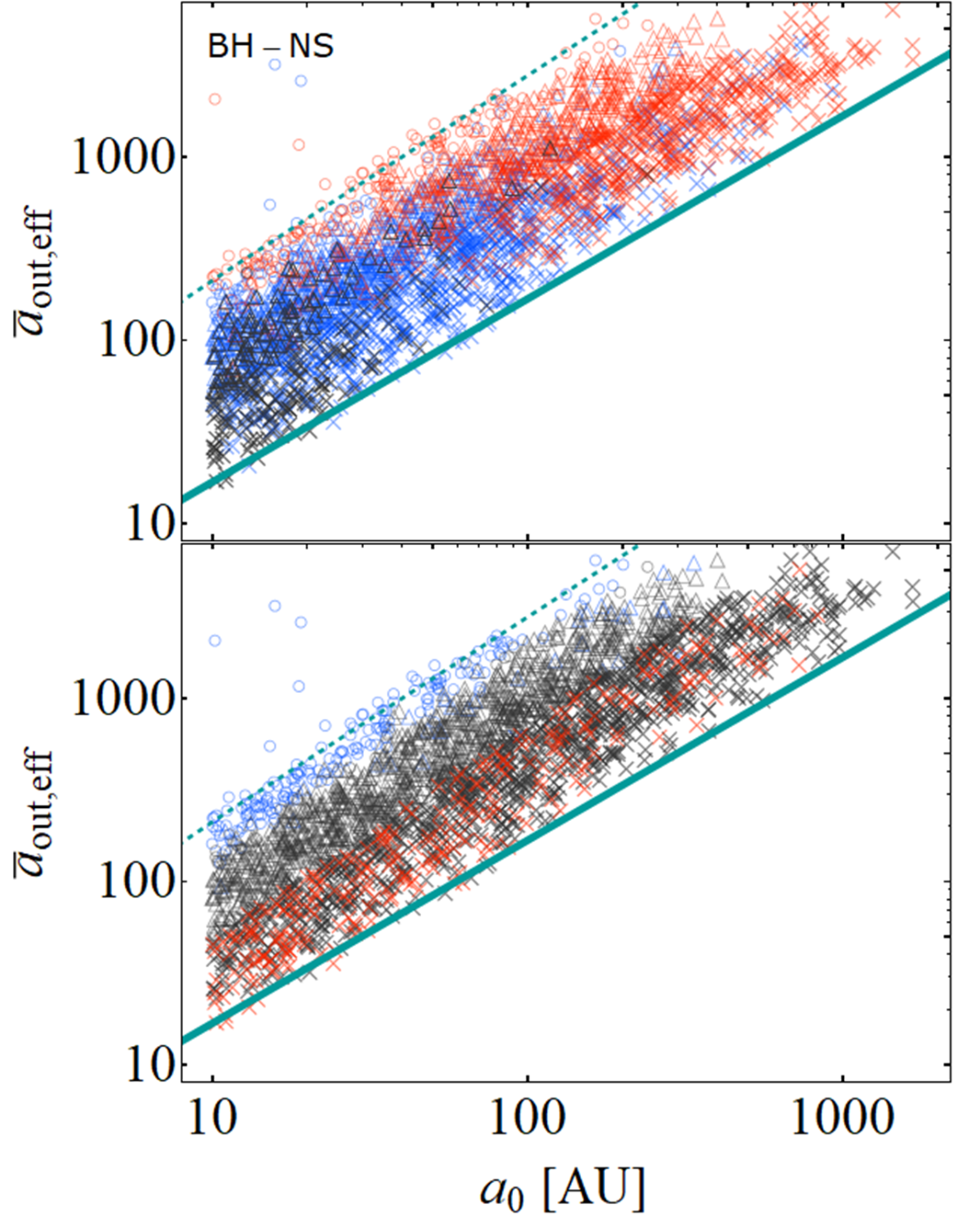}
\end{tabular}
\caption{Same as Figure \ref{fig:BH-BH a0}, but for our canonical BH-NS binaries.
}
\label{fig:BH-NS a0}
\end{figure}

As shown in Figure \ref{fig:BH-BH I0}, there is a preference for initially highly inclined systems (around $I_0\sim90^\circ$)
to generate mergers. This is the typical outcome for ``quadrupole" systems, for which the two members of the inner binary have comparable masses
such that $\varepsilon_\oct\ll 1$ (see Equation \ref{eq:epsilon oct}).
The distribution shows that mergers occur at a wide range of initial inclinations,
but the number decreases sharply when $I_0$ goes below $40^\circ$ or above $140^\circ$.
Therefore, we do not simulate the triples outside this ``Kozai window".
From the middle and bottom panels, we
see that the quick and slow mergers have no obvious trend in $a_\OUT/a_0$ and $\bar{a}_{\OUT,\eff}/a_0$.
However, the majority of mergers happen with $a_\OUT/a_0\lesssim100$ and $\bar{a}_{\OUT,\eff}/a_0\lesssim20$.

In Figure \ref{fig:BH-BH a0}, we plot the initial conditions in terms of ($\bar{a}_{\OUT,\eff}$, $a_0$)
for BH mergers presented in Figure \ref{fig:BH-BH I0}. The upper and lower panels show the outcomes indicated by
different merger times and ``residual" eccentricities (i.e., the binary eccentricity when the peak GW frequency enters the LIGO band).
The solid line comes from the stability criterion (Equation \ref{eq:stable}), set by $e_0=1$, $e_\OUT=0$
and $m_3=m_{12}=50M_\odot$. The dashed line is obtained by the condition
\be\label{eq:fitting formula}
T_{\mathrm{m},0}(1-e_\m^2)^3=10^{10}\mathrm{yrs},
\ee
where $T_{\mathrm{m},0}\equiv(5c^5 a_0^4)/(256 G^3 m_{12}^2 \mu)$ is the merger time due to GW radiation of an isolated binary with the
initial semi-major axis $a_0$ and eccentricity $e_0=0$, and $e_\m$ is evaluated by Equation (\ref{eq:ELIM}) at
$e_0=0$ and $\eta_0\rightarrow0$. In Equation (\ref{eq:fitting formula}), the left-hand side represents the merger time of the
inner binary when its eccentricity is excited to $e_\m$ during a LK cycle \citep[][]{Liu-ApJ,Xianyu merger time}
\footnote{Note that this expression is valid only for systems in the DA regime and with $\varepsilon_\oct\ll1$
and $1-e_\m\ll1$; see \citet{Liu-ApJL,Liu-ApJ}}.
Thus, Equation (\ref{eq:fitting formula}) provides the upper limit of merger time for detectable LK-induced mergers.

In the upper panel of Figure \ref{fig:BH-BH a0}, we see that the quick BH mergers ($T_\mathrm{m}\lesssim 10^6$ yrs) are likely to occur
when both $\bar{a}_{\OUT,\eff}$ and $a_0$ are relatively small.
In the lower panel, the eccentric mergers ($e_\mathrm{m}\gtrsim0.1$) preferentially arise for combinations of ($\bar{a}_{\OUT,\eff}$, $a_0$)
that are located near the stability boundary.
This implies that the highest eccentricity excitations require the strongest perturbers.

For our canonical BH-NS binaries (with $m_1=30M_\odot, m_2=1.4M_\odot$), the octupole effect becomes important due to the high mass ratio.
Previous works \citep[]{Liu et al 2015,Anderson et al 2017} have shown that the main effect of the octupole potential
is to broaden the range of the initial $I_0$ for extreme eccentricity excitations ($e_\m=e_\li$).
In our simulations, we find 2683 mergers out of 26238 stable triple systems, with a merger fraction of $10.2\%$; this includes
209 mergers out of 9761 systems in DA regime (fraction $\simeq 2.14\%$),
1197 mergers out of 6010 systems in SA regime (fraction $\simeq 20\%$) and 1277 out of 10467 systems
in direct NB simulations (fraction $\simeq 12\%$).
The somewhat smaller merger fraction in the NB regime compared to the SA regime arises because some systems
become dynamically unstable during the NB integrations.

Figure \ref{fig:BH-NS I0} shows the dependence on the initial parameters for BH-NS merger events.
Compared to Figure \ref{fig:BH-BH I0} we see that a large range of the initial inclinations inside the Kozai window produces mergers,
as a result of the non-negligible $\varepsilon_\oct$ for BH-NS systems.

Figure \ref{fig:BH-NS a0} shows the similar result as Figure \ref{fig:BH-BH a0}, but for BH-NS binaries.
Since the merger fraction is much larger than the BH-BH case, the statistical features (parameter spaces)
for producing quick or eccentric mergers become more evident.

\subsection{Merger Properties}
\label{sec 3 3}

\begin{figure*}
\centering
\begin{tabular}{cccc}
\includegraphics[width=5cm]{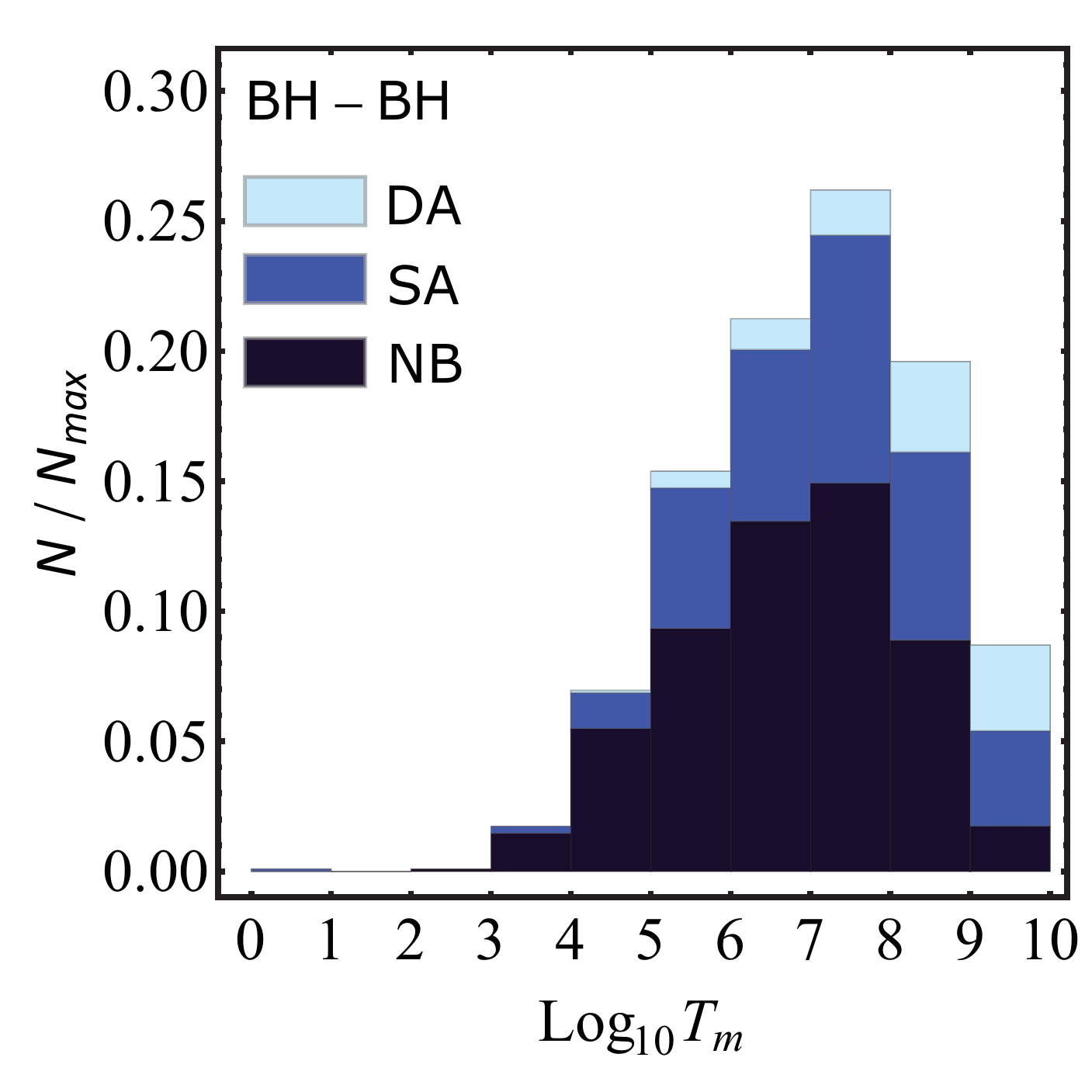}&
\includegraphics[width=5cm]{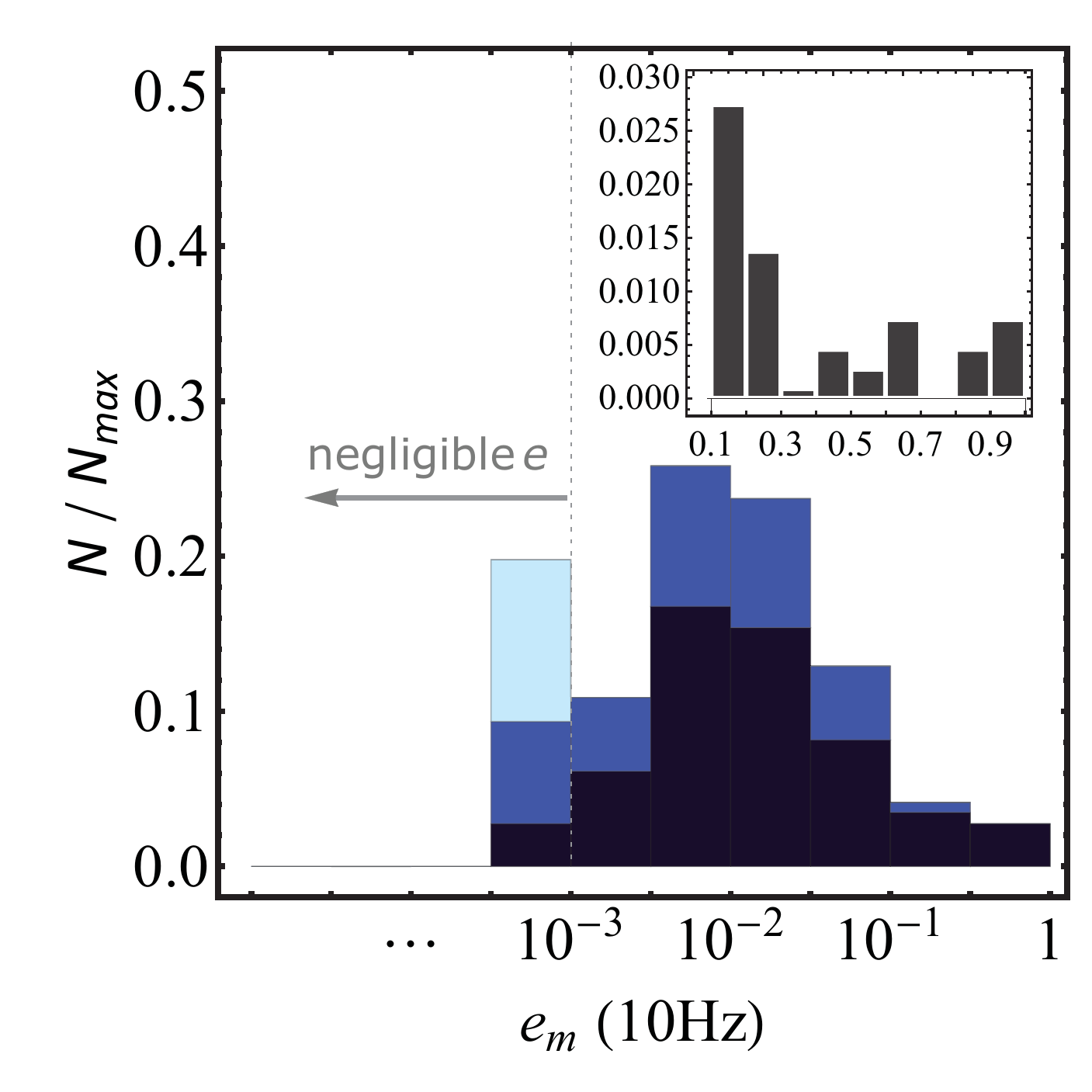}\\
\includegraphics[width=5cm]{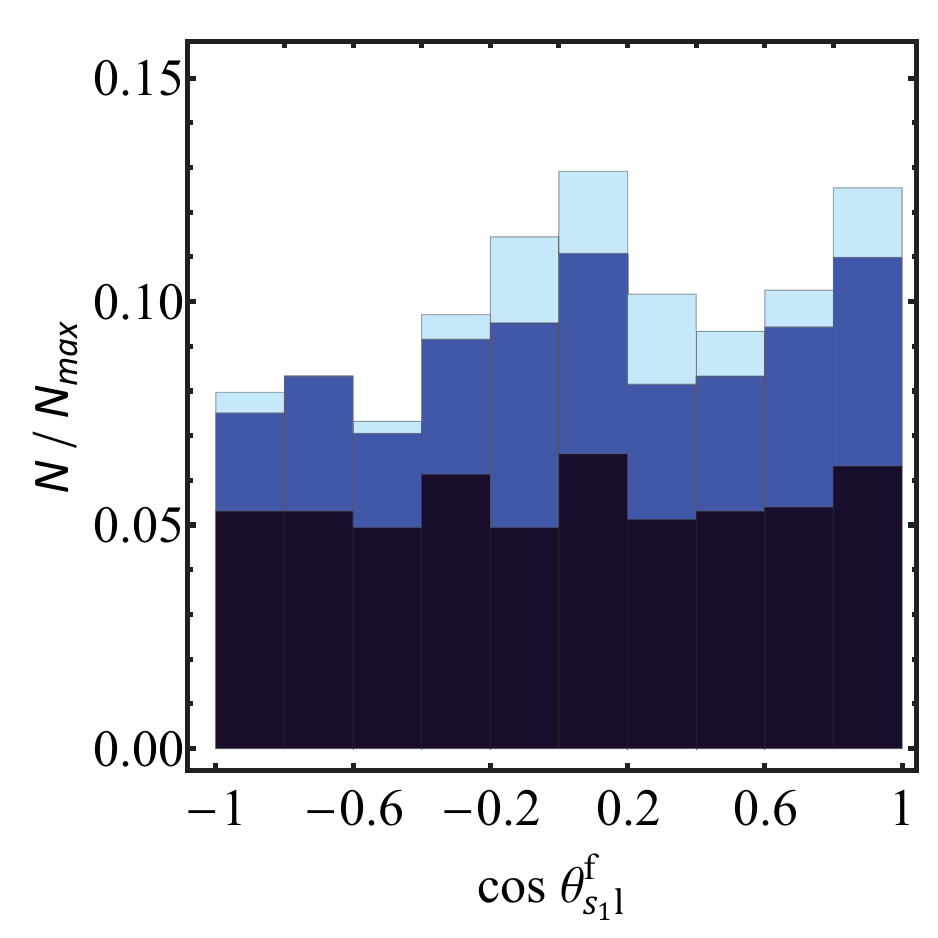}&
\includegraphics[width=5cm]{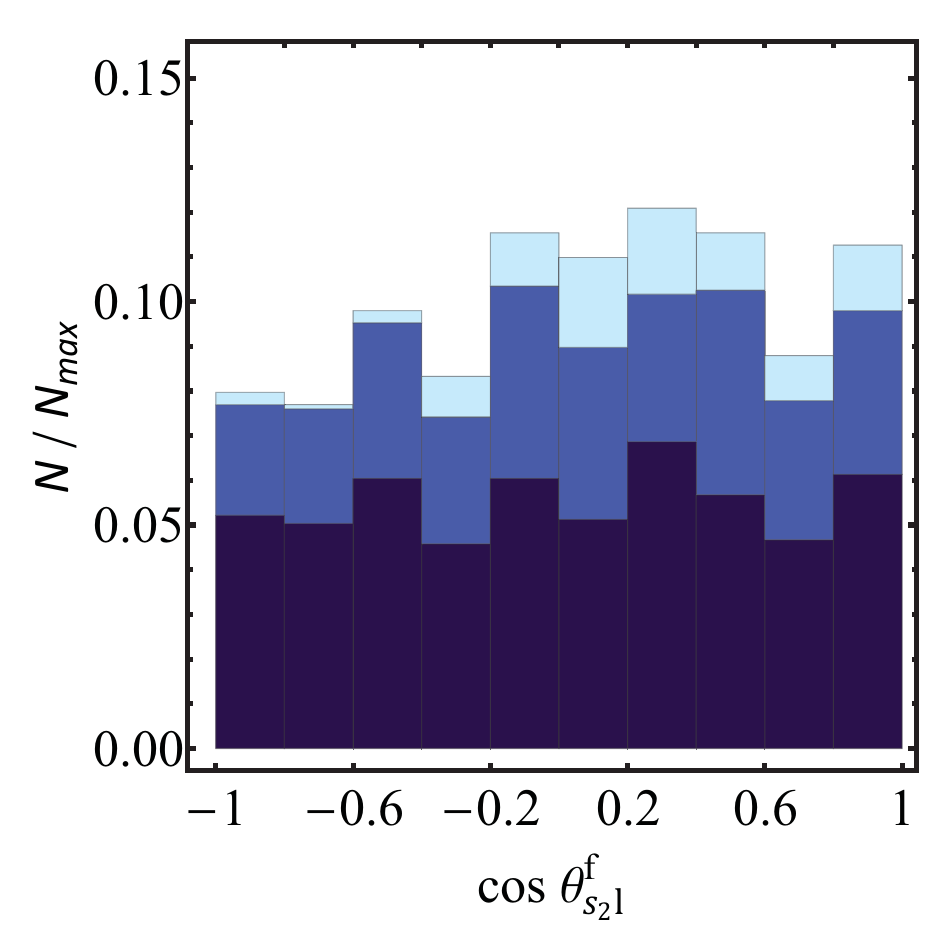}\\
\end{tabular}
\caption{Distributions of the merger time $T_\mathrm{m}$, eccentricity $e_\mathrm{m}$
(when the binary enters the LIGO band at 10 Hz), and spin-orbit misalignment angle for BH-BH mergers
in triples (see Figure \ref{fig:BH-BH I0}).
The results are for systems in different regimes, corresponding to different integration methods (DA, SA and NB).
All the data is normalized by the total number of mergers, $N_\m=1092$, including
114 (DA), 373 (SA) and 605 (NB) mergers out of 25255 simulated systems.
In the top right panel, we group all the $e_\mathrm{m}<10^{-3}$ mergers together and
show the zoom-in version of the mergers with $e_\mathrm{m}>0.1$.
}
\label{fig:BH-BH PS}
\end{figure*}

\begin{figure*}
\centering
\begin{tabular}{cccc}
\includegraphics[width=5cm]{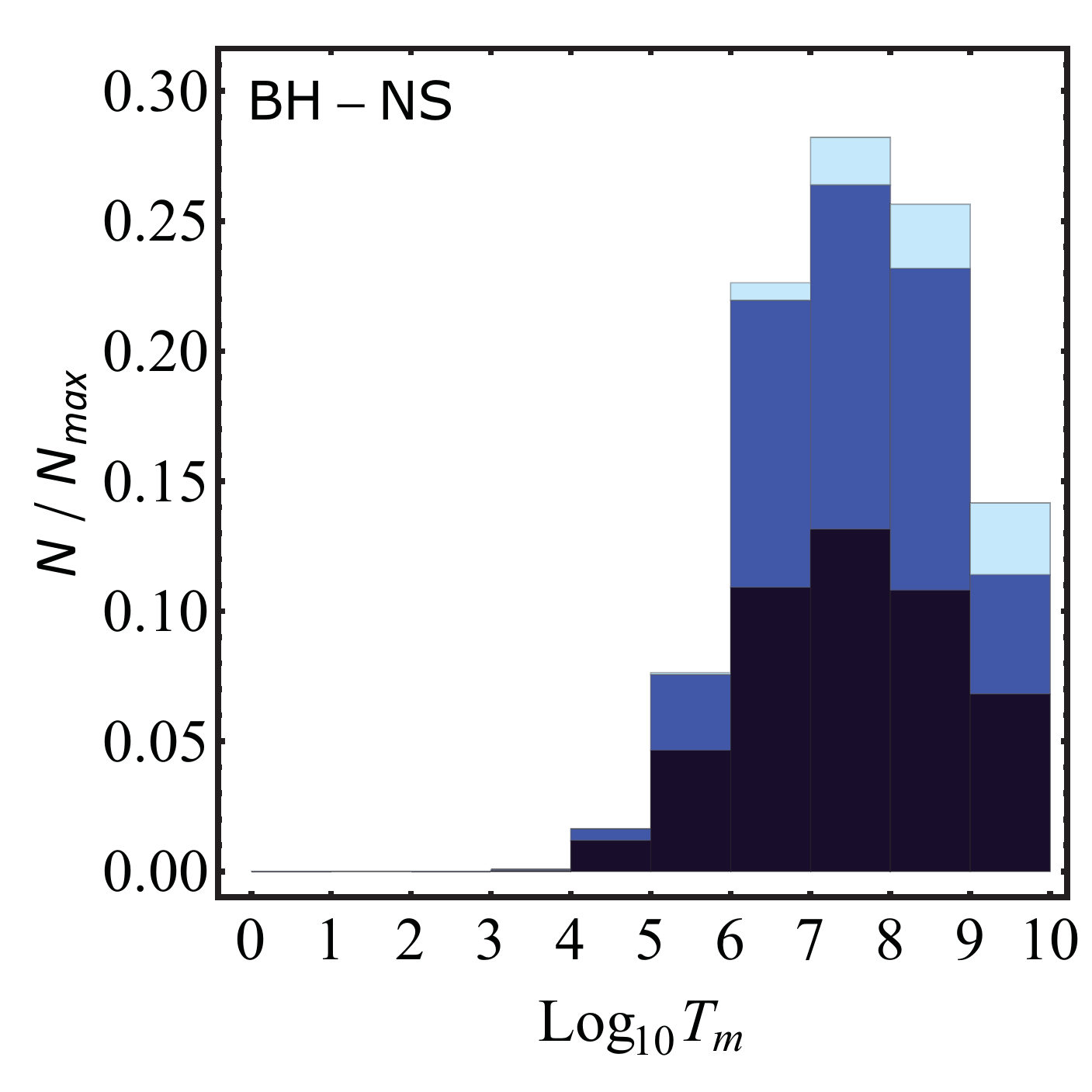}&
\includegraphics[width=5cm]{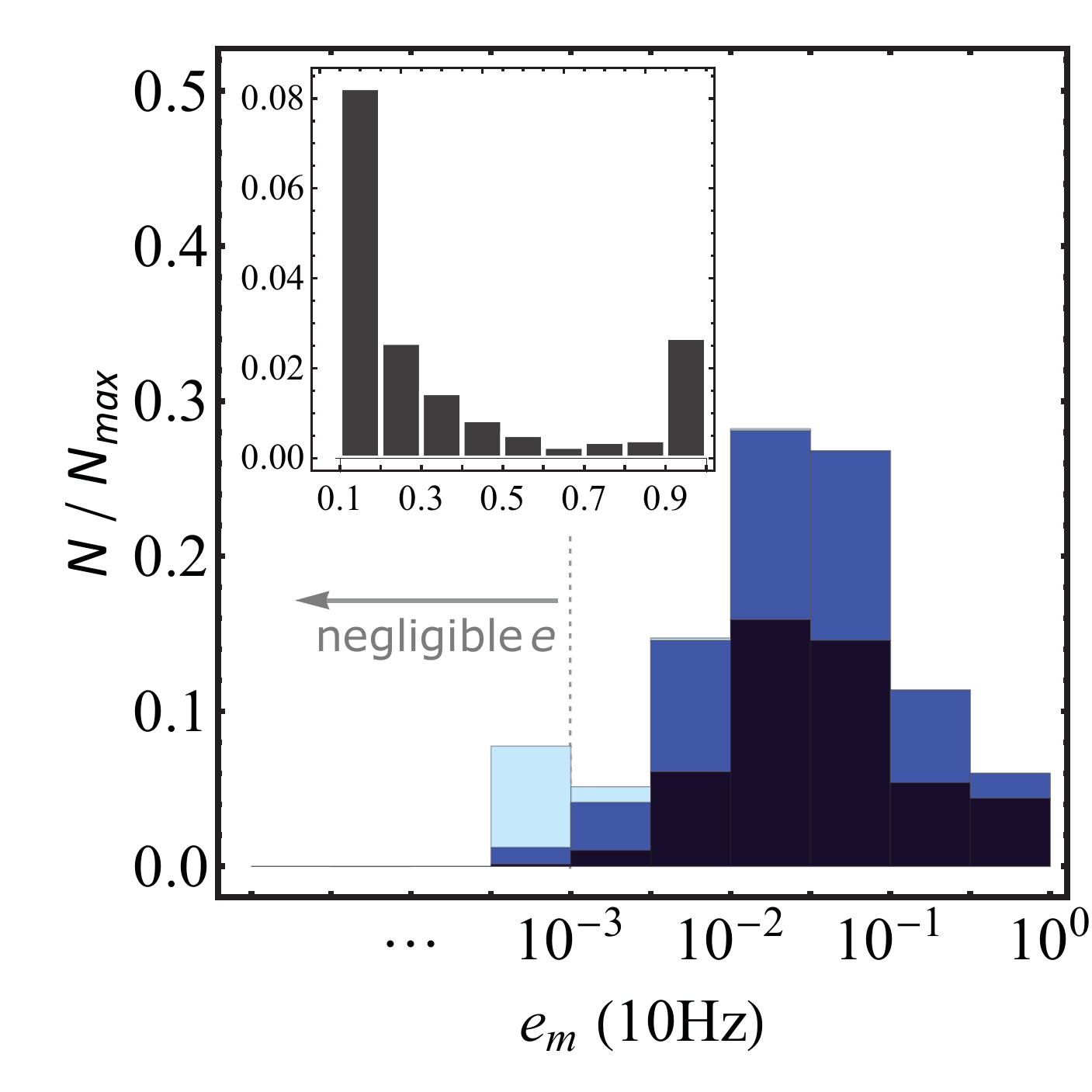}\\
\includegraphics[width=5cm]{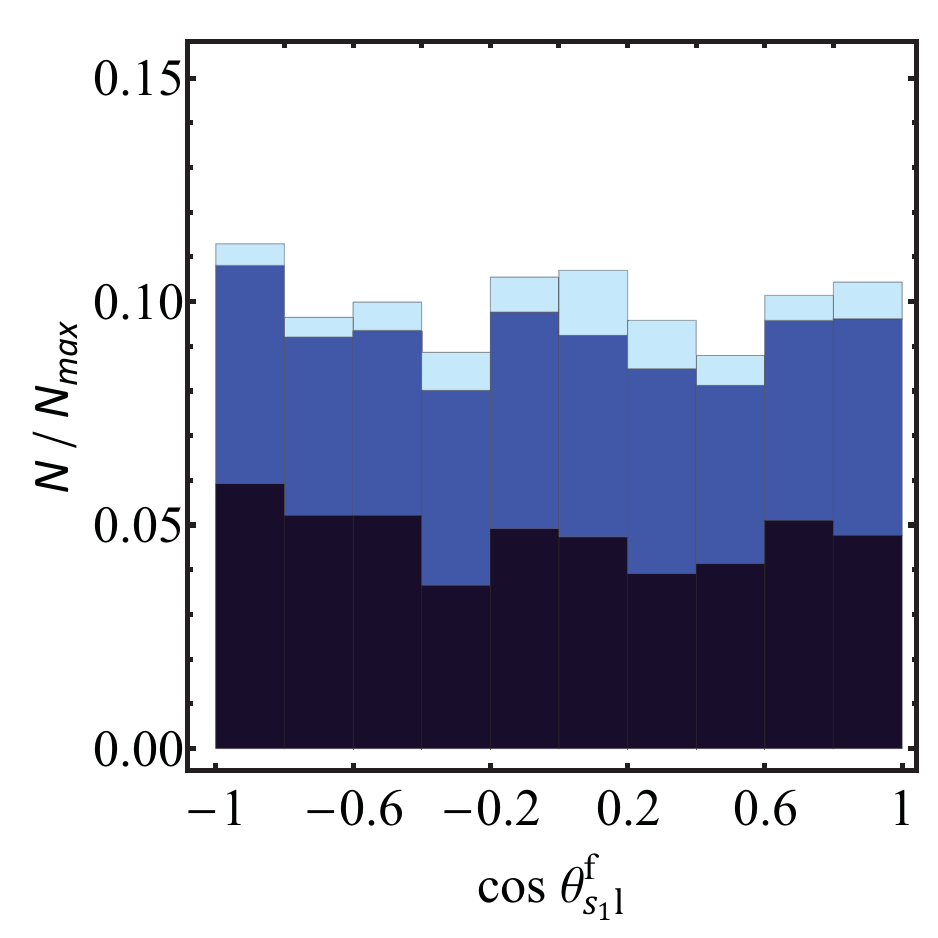}&
\includegraphics[width=5cm]{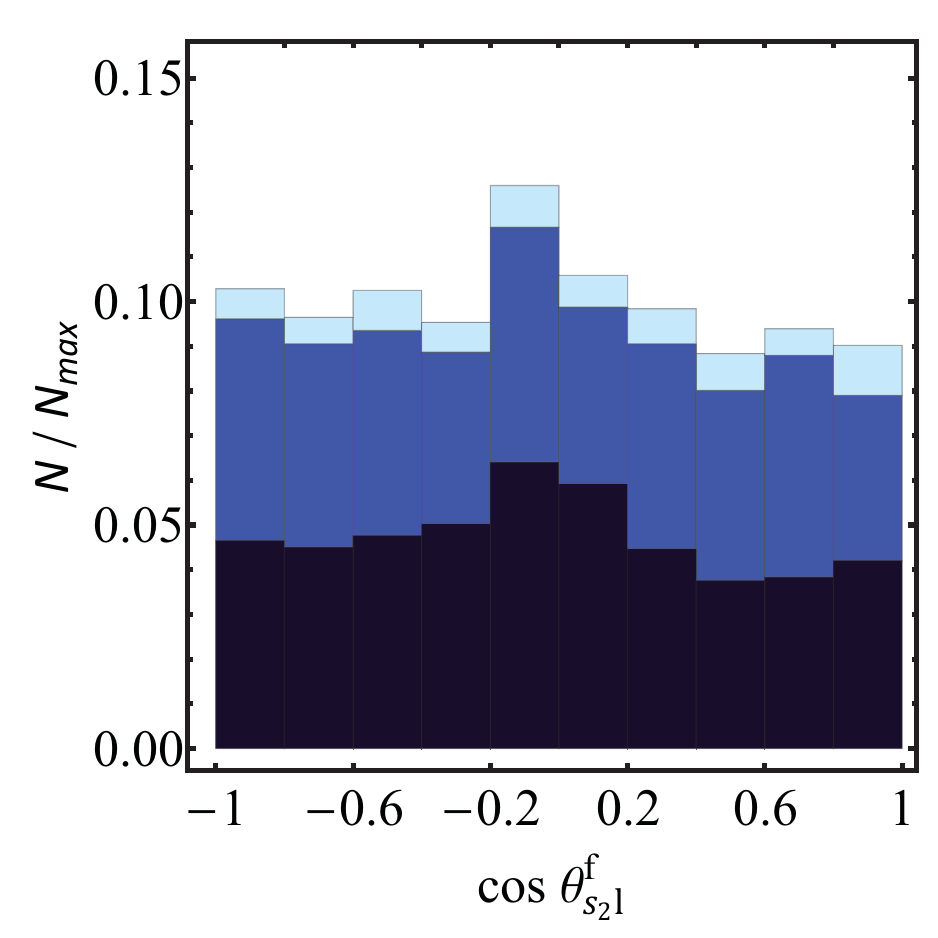}\\
\end{tabular}
\caption{Same as Figure \ref{fig:BH-BH PS}, except for BH-NS binaries (see Figure \ref{fig:BH-NS I0}).
The data represents $N_\m=2683$ mergers, including 209 (DA), 1197 (SA) and 1277 (NB) mergers out of 26238 simulated systems.
}
\label{fig:BH-NS PS}
\end{figure*}

\begin{figure}
\centering
\begin{tabular}{cccc}
\includegraphics[width=4cm]{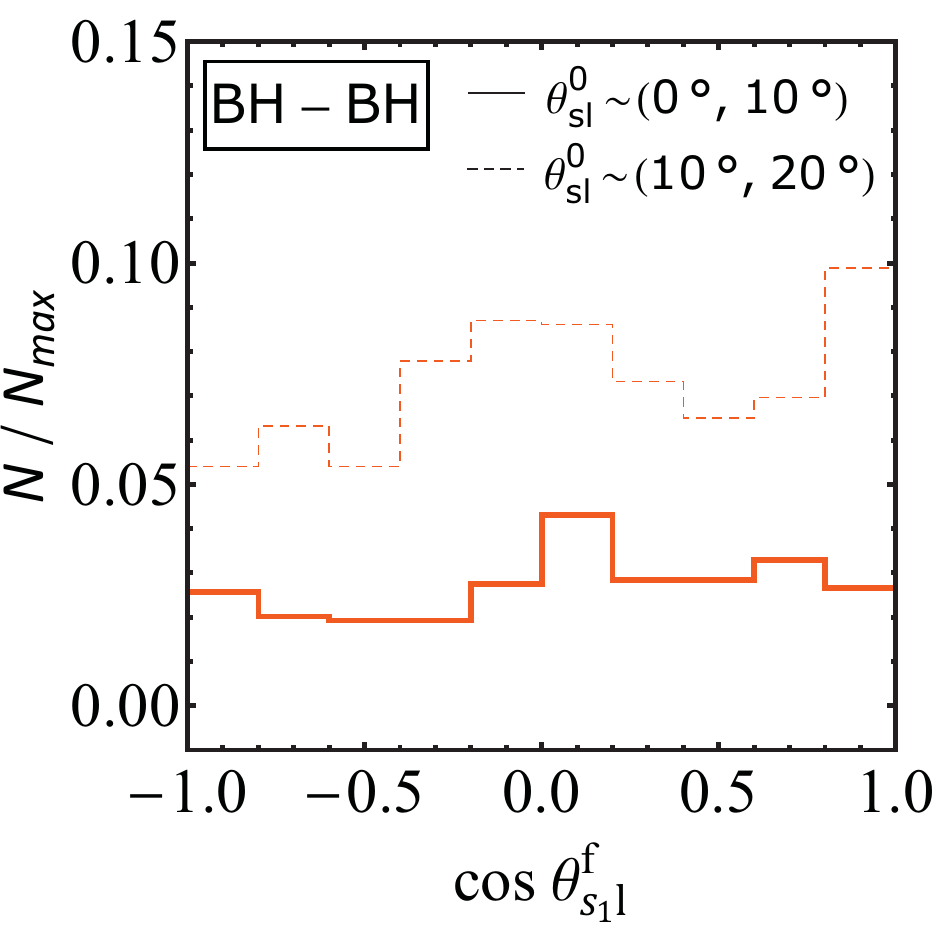}
\includegraphics[width=4cm]{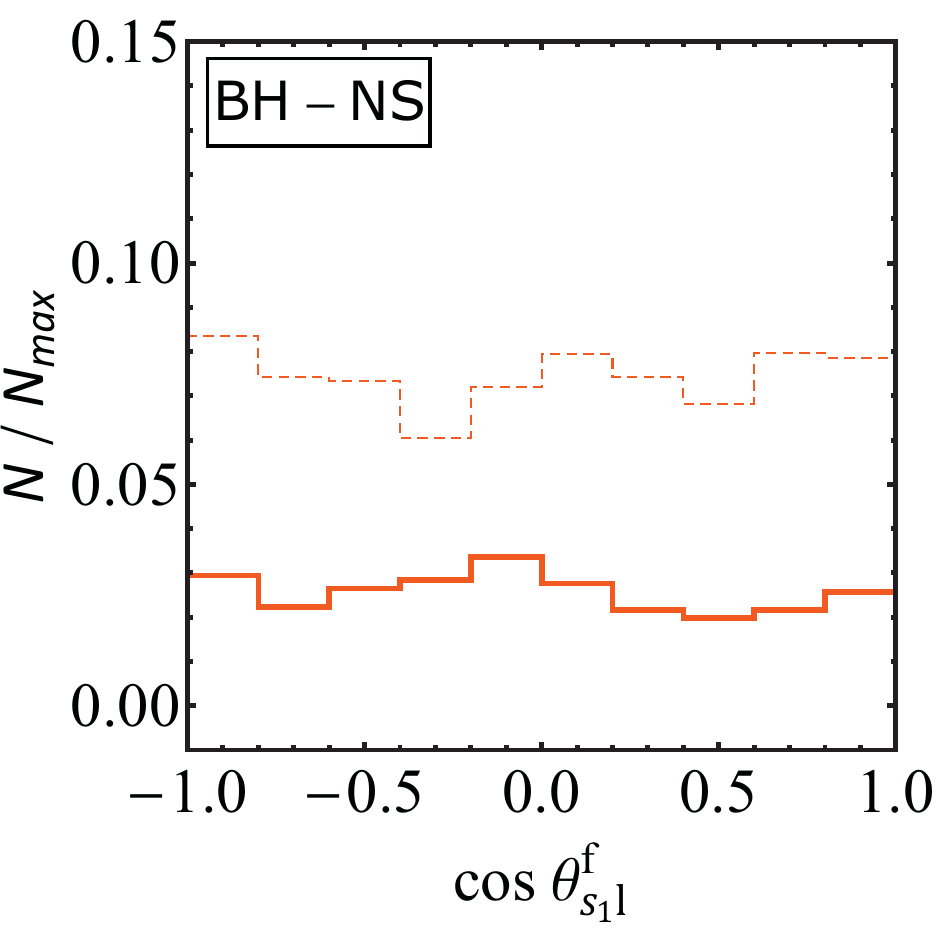}
\end{tabular}
\caption{Distribution of the final spin-orbit misalignment $\theta_\mathrm{s_1l}^\f$ for BH-BH mergers and BH-NS mergers in triples.
We separate the mergers with different ranges of
initial spin-orbit orientations, i.e., $\theta_\SL^0\in(0^\circ, 10^\circ$) and $\theta_\SL^0 \in(10^\circ, 20^\circ$).
}
\label{fig:diff initial spin}
\end{figure}

In Figure \ref{fig:BH-BH PS}, we show the distributions of the merger time $T_\mathrm{m}$, merger eccentricity $e_\mathrm{m}$ and
spin-orbit misalignment angle $\theta_\SL^\f$
for our canonical BH-BH mergers.
As shown in the top left panel, most merger events occur around $T_\mathrm{m}\sim 10^{7-8}$ yrs.
We find that about $80\%$ of the binaries enter the LIGO band with
$e_\mathrm{m}\gtrsim10^{-3}$, $50\%$ have $e_\mathrm{m}\gtrsim10^{-2}$, $7\%$ have $e_\mathrm{m}>0.1$, and $0.7\%$ have $e_\mathrm{m}>0.9$
(see the top right panel).
In the lower panels, we see that there is a slight clustering around $90^\circ$ for the final
spin-orbit misalignment $\theta_\SL^\f$.
This $90^\circ$ ``attractor" arise from quadruple-dominated mergers [corresponding to CASE I discussed in Section 4.2 in \citet{Liu-ApJ}].
The other peak around $\theta_\SL^\f=0^\circ$
is associated with CASE III and IV [see \citet{Liu-ApJ} for details].

Figure \ref{fig:BH-NS PS} shows the results for our canonical BH-NS binaries.
Because of the low NS mass,
the orbital decay due to GW is not as efficient as the BH-BH case.
As seen in the upper left panel, there is only a few merger events with $T_\mathrm{m}<10^5$ yrs.
Compared to the BH-BH case,
the high mass ratio of BH-NS binaries leads to stronger octupole effect,
which enhances extreme eccentricity excitation. In the upper right panel,
we see a significant increase in the number of mergers with appreciable $e_\mathrm{m}$:
Out of all merger events, about $93\%$ have $e_\mathrm{m}\gtrsim10^{-3}$, $80\%$ have $e_\mathrm{m}\gtrsim10^{-2}$,
$18\%$ have $e_\mathrm{m}>0.1$,
and $2.5\%$ have $e_\mathrm{m}>0.9$.
The octupole effect may also produce chaotic orbital evolution \citep[e.g.,][]{Lithwick 2011,Li chaos,Liu et al 2015},
such that the spin-orbit misalignment angle is allowed to settle to any value
\citep[CASE II in][]{Liu-ApJ}.
Thus, we find in the lower panels of Figure \ref{fig:BH-NS PS} that the final spin-orbit misalignments are largely
isotropic (i.e., uniform in $\cos \theta_\SL^\f$).

To determine how the final spin-orbit misalignment $\theta_\SL^\f$ depends on the initial $\theta_\SL^0$,
Figure \ref{fig:diff initial spin} shows the distribution of $\theta_\SL^\f$ for different ranges of initial $\theta_\SL^0$
($0^\circ-10^\circ$ versus $10^\circ-20^\circ$). We see that for BH-BH mergers, the peak around $\theta_\SL^\f=90^\circ$
exist regardless of the initial range of $\theta_\SL^0$. For BH-NS mergers, the distribution of $\cos\theta_\SL^\f$
is approximately uniformly for different ranges of $\theta_\SL^0$.

Several previous works also studied eccentric mergers of BH binaries
induced by a tertiary companion \citep[][]{Antonini 2017} \citep[see also][]{Rodriguez Spin} combined stellar evolution (mass loss and
natal kick) and dynamics in isolated triples, and found that about
$10\%$ of BH mergers have $e_\mathrm{m} > 0.1$, similar to our result.  On the
other hand, \citet{Silsbee and Tremaine 2017} also considered stellar
evolution and used N-body integration to evolve the triples, and found
that a few percent of the mergers may have $e_\mathrm{m} > 0.999$ (In
contrast we found about $1\%$ having $e_\mathrm{m}\gtrsim 0.9$). In the case
of BH mergers induced by a massive BH, \citet{Antonini 2012}
claimed that $10\%$ have $e_\mathrm{m}>0.1$ \citep[see also][]{Fragione GC IMBH}.
In dense star clusters, about $1\%$ of the BH mergers generated by
scattering processes may have $e_\mathrm{m}>0.1$ \citep[e.g.,][]{Samsing 2017,Rodriguez 2018}.
Thus, despite the uncertainties in
various scenarios, the detection of eccentric BH mergers would
constrain the dynamical formation channels of binary BHs.

\subsection{$\chi_\eff$ and Correlation with Merger Eccentricity $e_\mathrm{m}$}
\label{sec 3 4}

\begin{figure}
\centering
\begin{tabular}{cccc}
\includegraphics[width=8cm]{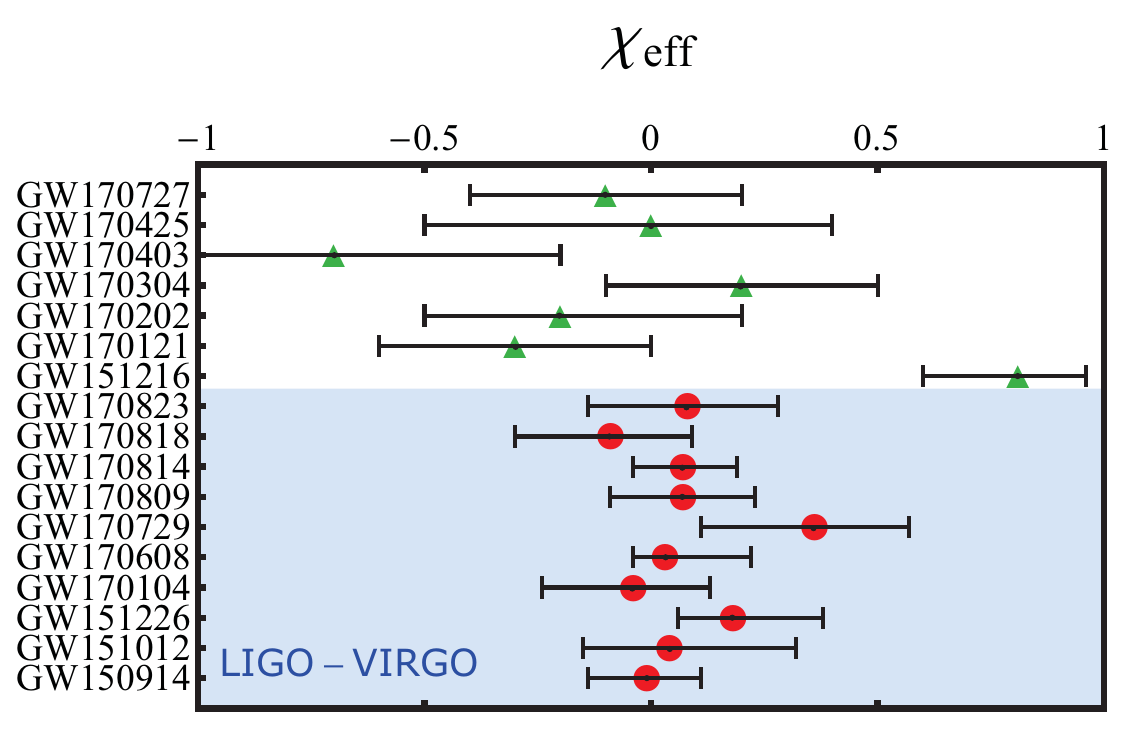}\\
\includegraphics[width=8cm]{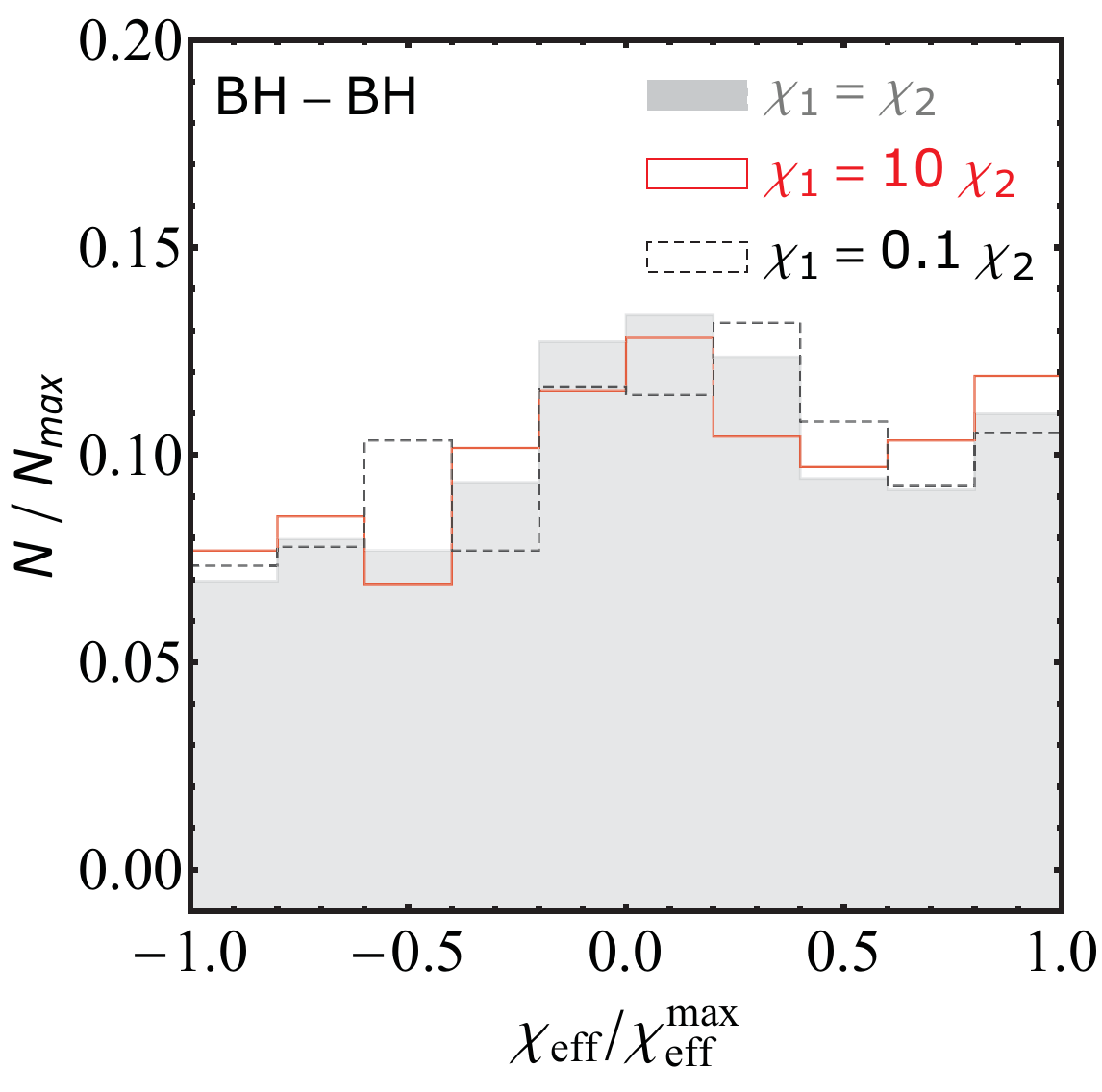}
\end{tabular}
\caption{Upper panel: All the public detection of BH-BH mergers (as of April 2019),
reported by \citet{Abbott 2018a} (red circle) and \citet{Zackay 2019,Venumadhav 2019}
(green triangle).
Lower panel: The distribution of the rescaled binary spin parameter
$\chi_\eff$ [Equations \ref{eq:chi eff}; and $\chi_\eff^\m=(m_1\chi_1+m_2\chi_2)/m_{12}$]
for the merging BH-BH binaries depicted in Figure \ref{fig:BH-BH PS}.
}
\label{fig:effective spin}
\end{figure}

Having obtained the distributions of $\cos\theta_\mathrm{s_1l}^\f$ and $\cos\theta_\mathrm{s_2l}^\f$ in Section \ref{sec 3 3}, we can
calculate the distribution of the effective spin parameter
\be\label{eq:chi eff}
\chi_{\rm eff}={m_1 \chi_1\cos\theta_\mathrm{s_1l}^\f + m_2 \chi_2\cos\theta_\mathrm{s_2l}^\f\over m_{12}},
\ee
where $\chi_{1,2}\leqslant1$ are the Kerr parameters.
Figure \ref{fig:effective spin} (lower panel) shows three examples with different $\chi_{1,2}$.
Here, $\chi_\eff^\m=(m_1\chi_1+m_2\chi_2)/m_{12}$ is the
maximum possible value of $\chi_\eff$ for given $m_1\chi_1$ and $m_2\chi_2$
(this maximum is achieved at $\cos\theta_\mathrm{s_1l}^\f=\cos\theta_\mathrm{s_2l}^\f=1$).
The different ratios of $\chi_1$ and $\chi_2$ affect the distribution of $\chi_\eff/\chi_\eff^\m$, but not significantly.
The peak around $\chi_\eff\simeq0$ is clearly visible, although not as distinct as in \citet{Liu-ApJ},
who considered a more limited parameter space. The full range of $\chi_\eff$ values (from negative to positive) becomes possible.
This is in contrast to the isolated stellar binary evolution channel, which always predict positive $\chi_\eff$
because of the preferentially aligned spins after the binaries undergo mass transfer or tidal coupling
\citep[e.g.,][]{Zaldarriaga,Gerosa spin}.
Comparison with current LIGO/VIRGO detections (top panel of Figure \ref{fig:effective spin})
suggests that the triple-driven merger scenario may be required, but obviously it is premature to draw any
firm conclusion at this point, because of the (partial) degeneracy between $\chi_1$, $\chi_2$ and
spin-orbit misalignment angles.

For BH-NS binaries, $\chi_\eff\simeq\chi_1\cos\theta_\mathrm{s_1l}^\f$, and
the peak around $\chi_\eff\simeq0$ is insignificant because of the random distribution of the final spin-orbit
misalignment angles (see Figure \ref{fig:BH-NS PS}).

\begin{figure}
\centering
\begin{tabular}{cccc}
\includegraphics[width=8cm]{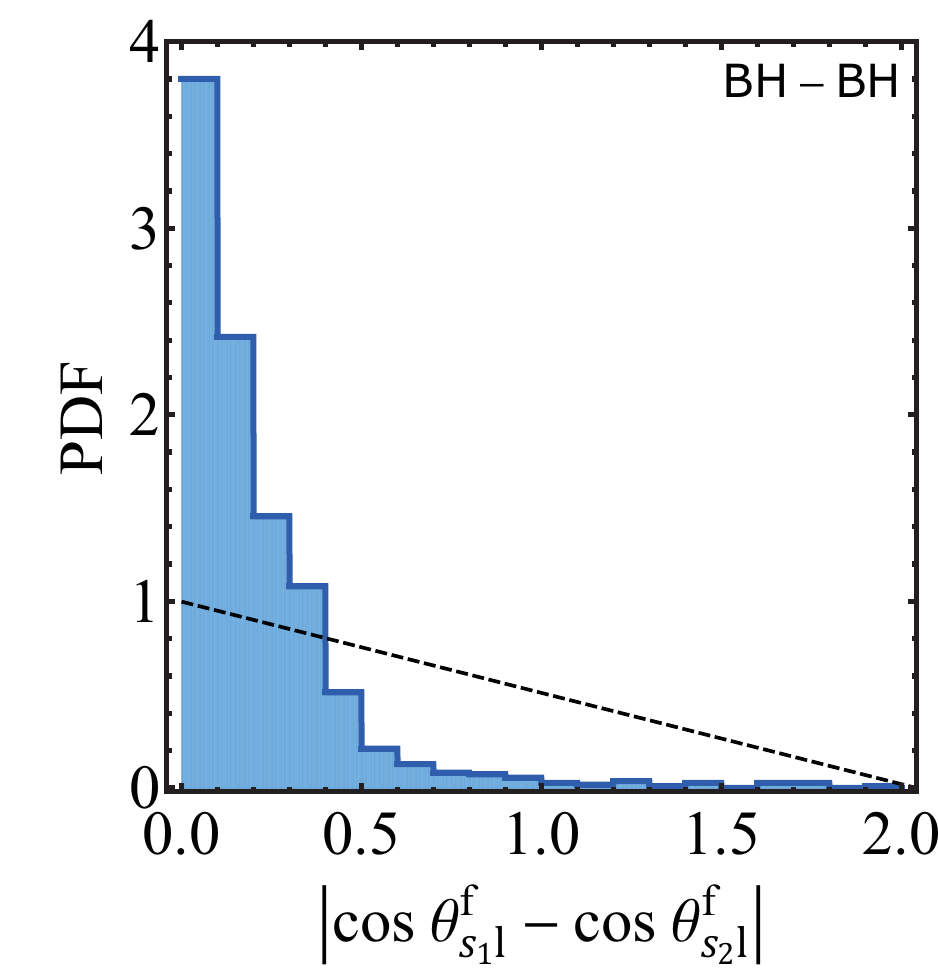}
\end{tabular}
\caption{The distribution of $|\cos\theta_\mathrm{s_1l}^\f-\cos\theta_\mathrm{s_2l}^\f|$ for merging BH-BH binaries,
where the data comes from Figure \ref{fig:BH-BH PS}.
The dashed line represents Equation (\ref{eq:p chi eff}), obtained assuming uncorrelated isotropic spin distributions.
}
\label{fig:spin correlation}
\end{figure}

\begin{figure*}
\centering
\begin{tabular}{cccc}
\includegraphics[width=17cm]{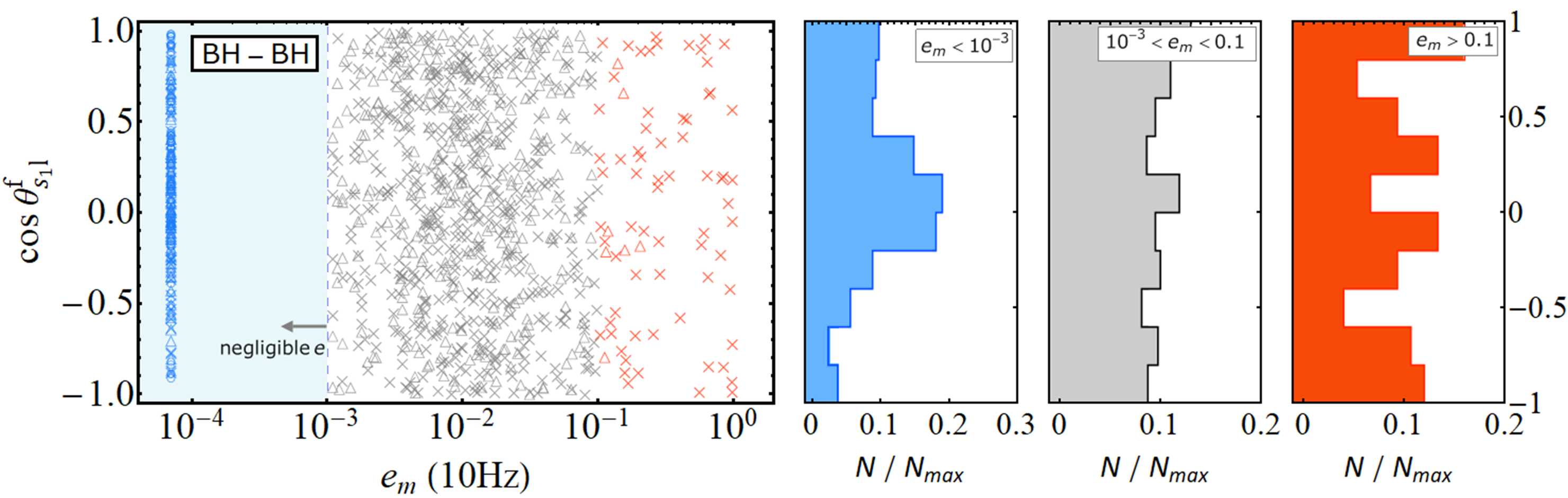}
\end{tabular}
\caption{Spin-orbit misalignment distribution for BH-BH systems with different merger eccentricities.
The left panel shows the final spin-orbit misalignment angles and
merger eccentricities of all BH-BH merger events in our simulations.
The color-coded symbols represent systems with different ranges of $e_\mathrm{m}$
(systems with $e_\mathrm{m}<10^{-3}$ are grouped together),
where the symbols indicate the mergers achieved by DA, SA and NB integrations (same as Figure \ref{fig:BH-BH I0}).
The right three panels show the distributions of $\cos\theta_\mathrm{s_1l}^\f$ for mergers with different
ranges of $e_\mathrm{m}$.
}
\label{fig:final spin and eccenticity BH}
\end{figure*}

\begin{figure*}
\centering
\begin{tabular}{cccc}
\includegraphics[width=17cm]{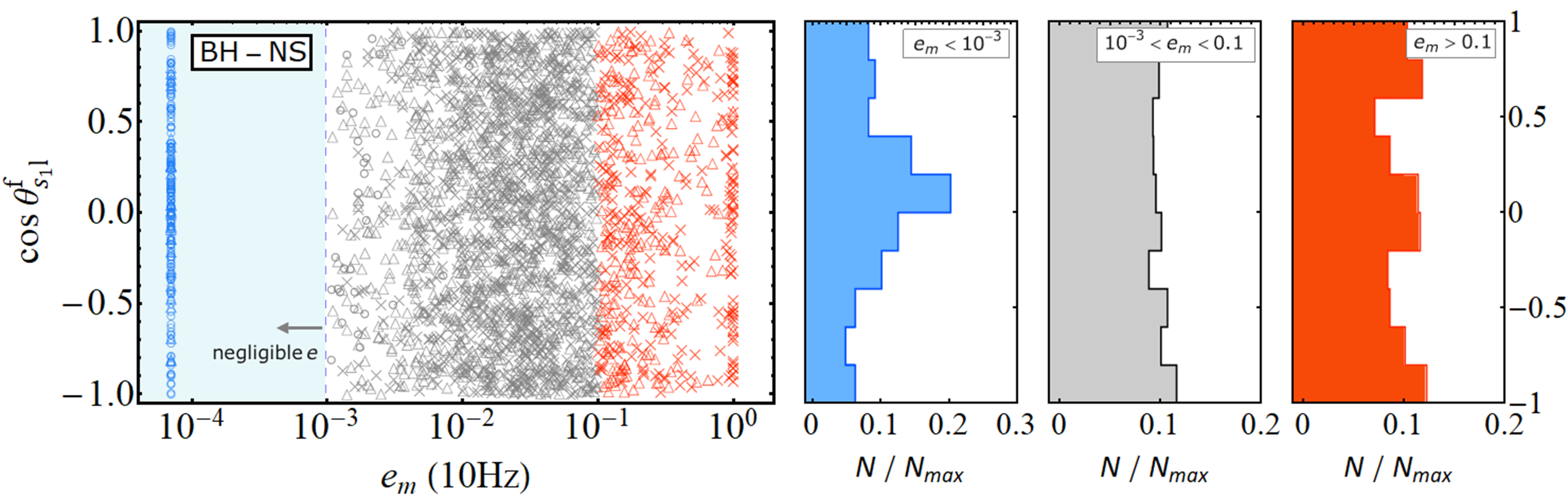}
\end{tabular}
\caption{Same as Figure \ref{fig:final spin and eccenticity BH}, but for BH-NS binaries.
}
\label{fig:final spin and eccenticity NS}
\end{figure*}

If the distributions of $\cos\theta_\mathrm{s_1l}^\f$ and $\cos\theta_\mathrm{s_2l}^\f$ are uncorrelated,
as we may expect to be the case for $m_1\neq m_2$, when the octupole effect is significant,
the distribution of $|\cos\theta_\mathrm{s_1l}^\f-\cos\theta_\mathrm{s_2l}^\f|$ can be
derived directly, where $\mu_1\equiv\cos\theta_\mathrm{s_1l}^\f$ and $\mu_2\equiv\cos\theta_\mathrm{s_2l}^\f$.
Given $P_1 (\mu_1)$ and $P_2 (\mu_2)$,
the distribution function of $\Delta\mu$ is
\be\label{eq:probability}
\begin{split}
P(\Delta\mu)=&\int^{1-\Delta\mu}_{-1}P_1(\mu_1)d\mu_1\int^{1-\Delta\mu}_{-1}P_2(\mu_2)d\mu_2\\
&\times\delta(\Delta\mu-\mu_1+\mu_2).
\end{split}
\ee
In the case where $\mu_1$ and $\mu_2$ are uniformly distributed, we have $P_1=P_2=1/2$,
and Equation (\ref{eq:probability}) gives
\be\label{eq:p chi eff}
P(\Delta\mu)=\left\{
\begin{array}{ccc}
\begin{split}
&(1/2)-(1/4)\Delta\mu,~~~~~\Delta\mu>0,\\
&(1/2)+(1/4)\Delta\mu,~~~~~\Delta\mu<0.
\end{split}
\end{array}
\right.
\ee
Figure \ref{fig:spin correlation} shows the distribution of $|\cos\theta_\mathrm{s_1l}^\f-\cos\theta_\mathrm{s_2l}^\f|$ from the
simulation data depicted in Figure \ref{fig:BH-BH PS}.
Compared to Equation (\ref{eq:p chi eff}), we see that the actual distribution is much more peaked around $\Delta\mu=0$.
This suggests that the two spins in the merging BH binary are correlated. Namely, in the majority of mergers,
the two BHs are likely to have similar final
$\theta_\SL^\f$, although each $\theta_\SL^\f$ could be any value running from $0^\circ$ to $180^\circ$.

Our previous work \citep[][]{Liu-ApJ} and those by \citet{Antonini spin} and \citet{Rodriguez Spin} found a distinct peak in
the $\chi_{\rm eff}$ distribution around $\chi_{\rm eff}=0$ for
sufficiently hierarchical systems. In this study, we have considered
all possible stable triples covering a wide levels of hierarchy. From
Figure \ref{fig:BH-BH PS}, we see that highly hierarchical systems only contribute a
small fraction of mergers, i.e. the majority of merging BH binaries
are produced by the moderately hierarchical systems with strong
tertiary companions. The resultant large number of ``isotropic" final
spin-orbit misalignment angles tend to ``bury" the $90^\circ$ signature. Thus,
the $\chi_{\rm eff}=0$ peak in the overall distribution shown in
Figure \ref{fig:effective spin} is less distinct than the one found in \citet{Liu-ApJ}
and \citet{Antonini spin}.

Figure \ref{fig:final spin and eccenticity BH} illustrates the correlation between the final spin-orbit misalignment angles and
merger eccentricities of BH-BH binaries.
We see that the systems with $e_\mathrm{m}\lesssim10^{-3}$ exhibit a peak around $\cos \theta_\SL^\f\simeq0$, while
those with larger $e_\mathrm{m}$ exhibit more uniform distribution in $\cos \theta_\SL^\f$. This
is consistent with the $90^\circ$ ``attractor" found in \cite{Liu-ApJ} (CASE I in that paper),
where we used DA secular equations to evolve the triple systems and spins.
In that work, no significant $e_\mathrm{m}$ was generated for the systems considered.

In Figure \ref{fig:final spin and eccenticity NS}, we consider BH-NS binaries.
As in the BH-BH case (Figure \ref{fig:final spin and eccenticity BH}),
systems with $e_\mathrm{m}\ll1$ exhibit a $90^\circ$ peak in the $\theta_\SL^\f$ distribution,
while those with large $e_\mathrm{m}$ do not.

\begin{figure}
\centering
\begin{tabular}{cccc}
\includegraphics[width=8cm]{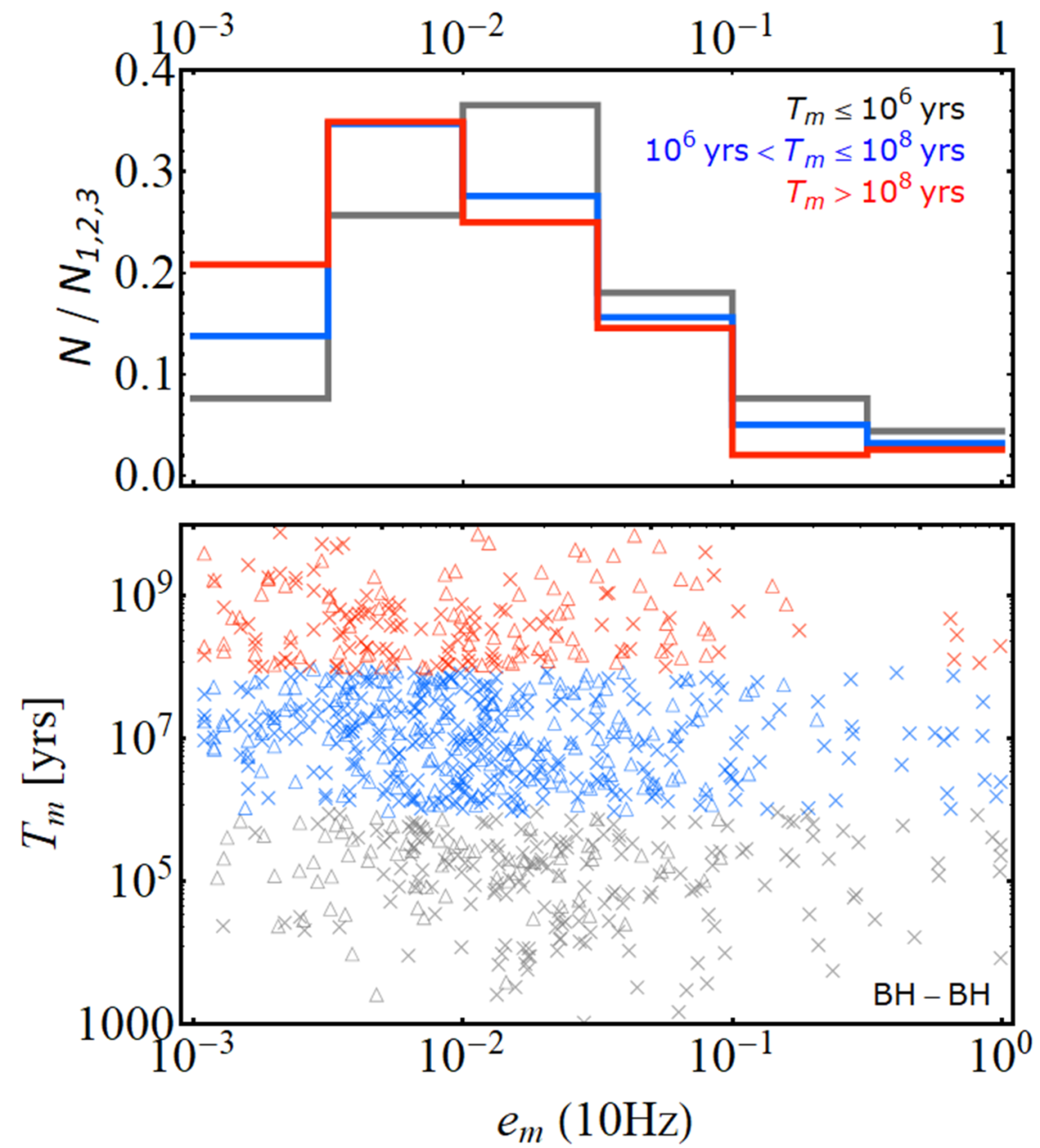}
\end{tabular}
\caption{Upper panel: the distribution of the residual eccentricity for BH-BH mergers with different range of merger times (color-coded),
normalized by the number of mergers in each category of $T_\mathrm{m}$.
Lower panel: the correlation between the merger time and residual eccentricity for each merger.
The symbols indicate the mergers achieved by DA, SA and NB integrations (same as Figure \ref{fig:BH-BH I0}).
}
\label{fig:merger time BH}
\end{figure}

\begin{figure}
\centering
\begin{tabular}{cccc}
\includegraphics[width=8cm]{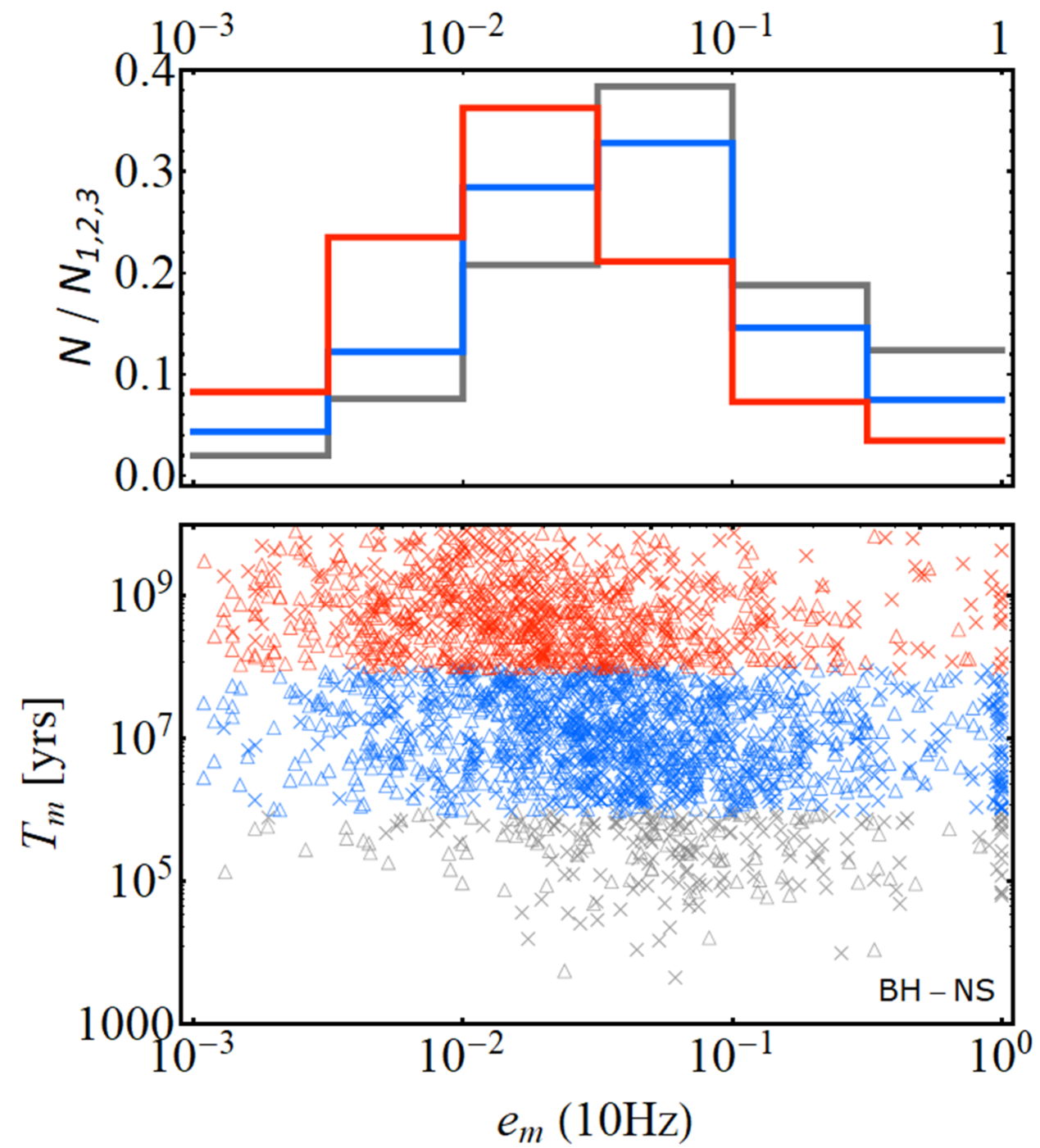}
\end{tabular}
\caption{Same as Figure \ref{fig:merger time BH}, but for BH-NS binaries.
}
\label{fig:merger time NS}
\end{figure}

In reality, triple systems (especially the outer orbit) can be perturbed (even disrupted)
by close fly-bys with other objects \citep[e.g.,][]{Perets flyby}.
The lifetime of a triple may be significantly shorten than $10^{10}$~years, depending on the stellar density of the
surroundings. To examine mergers before the disruption of triples,
we separate the mergers by the merger time ($T_\mathrm{m}\leq10^6 \mathrm{yrs}$,
$10^6 \mathrm{yrs}<T_\mathrm{m}\leq10^8 \mathrm{yrs}$ and $T_\mathrm{m}>10^8 \mathrm{yrs}$).
Figures \ref{fig:merger time BH} and \ref{fig:merger time NS} show that fast mergers (with shorter $T_\mathrm{m}$)
tend to have a
higher probability to be accompanied by eccentric orbits at 10 Hz (i.e., $e_\mathrm{m}>0.1$).
This is more evident for BH-NS binaries (Figure \ref{fig:merger time NS}),
where the octupole terms contribute to a large set of systems with high-$e$
excitation and shorter merger time.

\section{Summary and Discussion}
\label{sec 4}
\subsection{Summary of Key Results}
\label{sec 4 1}

In this paper, we have systematically studied the dynamical signatures
of BH-BH and BH-NS mergers induced by tertiary companions in triple
systems, emphasizing the detectable merger eccentricities and
spin-orbit misalignments when the inner binary enters the LIGO sensitivity
band ($>10$~Hz). Going beyond our previous works \citep[][]{Liu-ApJL,Liu-ApJ},
we examined a wide range of triple systems to explore the
dependence of the merger properties on the binary/triple
parameters. More specifically, for two types of binaries, with
$(m_1,m_2)=(30M_\odot,20M_\odot)$ (representing a canonical BH-BH
binary) and $(30M_\odot,1.4M_\odot)$ (representing a canonical BH-NS
binary), we considered all possible binary/triple configurations and
parameters that lead to binary mergers and determined the distributions
of merger times, eccentricities and
spin-orbit misalignments. We used both single-averaged and double-averaged secular
equations that include octupole terms and spin-orbit coupling \citep[already
presented in][]{Liu-ApJ}, as well as a newly developed N-body code based
on {\tt ARCHAIN} algorithm, to evolve systems with various degrees of hierachy.
This allowed us to efficiently cover a wide range of parameter space and
to determine the merger properties reliably.

We first explored what kind of initial triples can produce BH-BH and
BH-NS mergers within $10^{10}$~years. In addition to significant
tertiary inclinations (as required for the Lidov-Kozai effect to be
efficient), tertiary-induced mergers also require that the initial
tertiary-binary separation ratio lie in the range $5\lesssim
a_\OUT/a_0 \lesssim 100$, or more generally, the ratio between the
scaled outer semimajor axis to the inner one,
$\bar{a}_{\OUT,\eff}/a_0$ (see Eq. \ref{eq:aout bar}), ranges from 1 to 20 (see
Figs. \ref{fig:BH-BH I0} and \ref{fig:BH-NS I0}).  The stability
criterion of the triple and an analytical ``limiting'' merger time expression
(Eq. \ref{eq:fitting formula}) provide an excellent characterization of the parameter space
leading to mergers (see Figs. \ref{fig:BH-BH a0} and \ref{fig:BH-NS a0}).

Our studies have revealed several distinct dynamical signatures of the
tertiary-driven binary merger scenario. For merging binaries with
comparable masses (i.e., BH-BH binaries), we found that about $7 \%$
of the mergers have eccentricities $e_\mathrm{m}>0.1$ at 10
Hz, and $0.7\%$ have $e_\mathrm{m}>0.9$. Distant tertiary companions (with
negligible octupole effects; see Eq. \ref{eq:epsilon oct}) tend to
generate spin-orbit misalignments $\theta_\SL^\f$ around $90^\circ$
and negligible $e_\mathrm{m}$ (see
Figs. \ref{fig:BH-BH PS} and \ref{fig:final spin and eccenticity BH}).
Closer tertiary companions (with stronger octupole effects) produce a
more isotropic distribution of $\theta_\SL^\f$ and non-negligible
$e_\mathrm{m}$. Note that the misalignment angles $\theta_\SL^\f$ of
the two BHs are correlated (see Fig. \ref{fig:spin correlation}).  We
have also computed the distribution of the mass-weighted spin
parameter $\chi_{\rm eff}$ (Eq. \ref{eq:chi eff}) of merging BH
binaries in triples.  Although $\chi_\eff$ can have a wide range of
values, there still exists a characteristic shape with peak around
$\chi_\eff\simeq 0$ in its distribution (see Fig. \ref{fig:effective
  spin}).  This could serve as an indicator of tertiary-induced binary
mergers.

For merging binaries with high mass ratio (e.g., BH-NS binaries), we
found a large fraction of systems with significant merger
eccentricities as a result of the strong octupole effects: About
$18\%$ have $e_\mathrm{m}>0.1$ and $2.5\%$ have $e_\mathrm{m}>0.9$.
Thus, the residual eccentricity at 10~Hz could indeed serve
as an indicator for such tertiary-driven binary mergers.
On the other hand, we found that the final spin-orbit misalignments have
an approximately isotropic distribution (with an insignificant $90^\circ$ peak),
except for the weak octupole systems that generate mergers with
negligible residual eccentricities
(see Figs. \ref{fig:BH-NS PS} and \ref{fig:final spin and eccenticity NS}).

Overall, our study showed that a combination of detections of
$e_\mathrm{m}$ and $\theta_\SL^\f$ (or $\chi_{\rm eff}$) from future
LIGO/VIRGO observations would provide key information to determine or
constrain the formation channels of merging BH-BH and BH-NS binaries.

\subsection{Discussion}
\label{sec 4 2}

Although in this paper we have focused on two types of binaries with
specific masses -- these masses can be measured from LIGO/VIRGO
observations, our survey of parameter space for the initial
binaries/triples was extensive.  The statistical properties of
tertiary-driven mergers found in this paper can be extended to other
merging binaries with various mass ratios.

We have found that a small fraction of merging binaries in triples
enter the LIGO band with appreciable eccentricities ($7\%$ of BH-BH
binaries and $18\%$ of BH-NS binaries have $e_\mathrm{m}>0.1$).  However,
almost all such binaries pass through the LISA band with high
eccentricities \citep[see also][]{Samsing 2018b,Kremer LISA,Fang 2019,Xianyu LISA,Hoang 2019}.
Thus, joint observations by LIGO/VIRGO and LISA, where the information
on the BH spin and eccentricity could be extracted, will be very
useful to constrain the formation mechanisms of merging binaries,
especially for the tertiary-driven scenario.

We have focused on triple systems in this paper. For quadrupole system,
binary-binary interactions can significantly enhance the binary
merger fraction \citep[e.g.,][]{Hamers and Lai 2017,Fang 2018,Hamers (2018),Liu-Quad,Fragione Quadruple}.
Although the occurrence rate of stellar quadruples is
smaller than that of stellar triples, dynamically induced BH mergers in quadruple systems
may be an important channel of producing BH mergers \citep[][]{Liu-Quad}.
A systematically study of the dynamical signatures of this channel is beyond the scope of this paper,
although we expect that many similar features found here may carry over.
As shown in \citet{Liu-Quad}, if the parameters of a quadruple system satisfies certain resonance criterion,
the inner binary eccentricity can be driven close to unity in a chaotic way,
leading to a similar behavior for $\theta_\SL^\f$ and $e_\mathrm{m}$ as in the cases of
triples with strong octupole effects (i.e., random $\cos\theta_\SL^\f$ distribution and
large $e_\mathrm{m}$).

\section{Acknowledgments}

We thank Bonan Pu for useful discussion.
This work is supported in part by the NSF grant AST-1715246 and NASA
grant NNX14AP31G.  BL is also supported in part by grants from NSFC
(No. 11703068 and No. 11661161012). This work made use of the High
Performance Computing Resource in the Core Facility for Advanced
Research Computing at Shanghai Astronomical Observatory.


\end{document}